\documentclass[%
 amsmath,
 aps,
prb,
twocolumn,
floatfix,
]{revtex4-2}

\usepackage{graphicx}
\usepackage{dcolumn}
\usepackage{bm}
\usepackage{hyperref}
\usepackage[mathlines]{lineno}

\hypersetup{
  colorlinks=true, linkcolor=blue, urlcolor=blue,
  pdftitle={the FSOS draft}, citecolor=blue}

\usepackage{booktabs,siunitx}

\usepackage{mathptmx}

\usepackage{diagbox}

\newcommand{\GWG}{\ensuremath{GW\Gamma^{(1)}}\,}
\newcommand{\GnWnG}{\ensuremath{G_\text{0}W_\text{0}\Gamma^{(1)}_0\,}}

\newcommand{\bfr}{ {\bf r}} 
\newcommand{\bfrp}{ {\bf r'}} 

\newcommand{\GnWn}{\ensuremath{G_\text{0}W_\text{0}}\,}

\begin{document}

\title{Assessing the {\GnWnG} approach: Beyond \GnWn with Hedin's full second-order self-energy contribution}

\author{Yanyong Wang}
\affiliation{CAS Key Laboratory of Quantum Information, University of Science and Technology 
of China, Hefei, Anhui 230026, China}
\author{Patrick Rinke}
\affiliation{Department of Applied Physics, Aalto University School of Science, 00076 Aalto, Finland}
\author{Xinguo Ren}
\email{renxg@iphy.ac.cn}
\affiliation{Beijing National Laboratory for Condensed Matter Physics, Institute of Physics, Chinese Academy of Sciences, Beijing 100190, China
}


\date{\today}

\begin{abstract}

We present a self-energy approach for quasiparticle energy calculations that goes beyond Hedin's $GW$ approximation by adding the full second-order self-energy (FSOS-$W$) contribution.
The FSOS-$W$ diagram involves two screened Coulomb interaction ($W$) lines and adding the FSOS-$W$ to the $GW$ self-energy can be interpreted as first-order vertex correction to $GW$ (\GWG). Our FSOS-$W$ implementation is based on the resolution-of-identity technique and exhibits better than $O(N^5)$ scaling with system size for small to medium-sized molecules. We then present one-shot \GWG (\GnWnG) benchmarks for the $GW$100 test set and a set of 24 acceptor molecules. For semilocal or hybrid density functional theory starting points, \GnWnG systematically outperforms \GnWn for the first vertical ionization potentials (vIPs) and electron affinities (vEAs) of both test sets. Finally, we demonstrate that a static FSOS-$W$ self-energy significantly underestimates the quasiparticle energies.

\end{abstract}
\maketitle

\section{\label{sec:intro} Introduction}
The one-electron properties of interacting many-electron systems can be conveniently described 
by the Green's function  -- also called the electron propagator -- of the system. In particular, the one-electron excitation energies, i.e., the energy change due to addition or removal of one electron, are given rigorously by the poles of the interacting Green's function $G$ \cite{Fetter/Walecka:1971,martin_reining_ceperley_2016}. Through Dyson's equation, the target interacting Green's function can be linked to an easily computable non-interacting reference Green's function $G_0$ via the self-energy $\Sigma$. Thus, the question of calculating the interacting Green's function $G$ of a many-electron system boils down to finding computationally affordable yet sufficiently accurate approximations for the self-energy $\Sigma$.

In Hedin's formulation of the interacting many-body problem \cite{Hedin:1965}, he casts the many-body Schr\"{o}dinger's equation into a set of five inter-dependent differential-integral equations. Hedin's equations offer an elegant framework for deriving and computing self-energy approximations. The most prominent example is the $GW$ approximation to the self-energy \cite{Hedin:1965}, in which  $\Sigma$ is given by the product of $G$ and the screened Coulomb interaction $W$. $GW$ emerges naturally from Hedin's equations as first-order approximation for the electron self-energy.  Since its first applications to real materials in the 1980s \cite{Hybertsen/Louie:1986,Godby/Schlueter/Sham:1986}, the $GW$ approximation has become the method of choice for band structure calculations of solids \cite{Hybertsen/Louie:1986,Godby/Schlueter/Sham:1986,Aryasetiawan/Gunnarsson:1998,Rinke/etal:2005,Schilfgaarde/Kotani/Faleev:2006,Shishkin/Marsman/Kresse:2007,Gatti/Bruneval/Olevano/Reining:2007,Friedrich/etal:2010,Jiang/Blaha:2016,Golze/Dvorak/Rinke:2019,Rangel/etal:2020,Ren/etal:2021}. 
More recently, it has also gained attention in chemistry and applications to molecules and clusters are on the rise \cite{Rostgaard/Jacobsen/Thygesen:2010,Blase/Attaccalite/Olevano:2011,Foerster/etal:2011,Blase_review:2012,Ren/etal:2012,Caruso/etal:2012,Bruneval/Marques:2013,Setten/etal:2013,Setten/etal:2015,Knight/etal:2016,Maggio/Kresse:2017,Lange/Berkelbach:2018,Golze/etal:2018,Sunila/etal:2021}. 

In practical $GW$ calculations, different operational procedures have been adopted. The most popular is the so-called $\GnWn$ scheme, in which the self-energy $\Sigma$ is constructed from a non-interacting Green's function $G_0$ and the corresponding screened Coulomb interaction $W_0$. $G_0$ is typically obtained from a mean-field calculation based on density functional theory (DFT). $\GnWn$ performs well for a large variety of applications but unfortunately depends on the input $G_0$, henceforth termed the starting point. To eliminate this starting point dependence, self-consistent $GW$ schemes of different flavors have been explored, including, e.g., eigenvalue-self-consistent $GW$ \cite{Hybertsen/Louie:1985,Golze/Dvorak/Rinke:2019}, Green-function-only partially self-consistent $GW$ (denoted $GW_0$) \cite{vonBarth/Holm:1996,Shishkin/Marsman/Kresse:2007}, quasiparticle self-consistent $GW$ \cite{Faleev/Schilfgaarde/Kotani:2004,Schilfgaarde/Kotani/Faleev:2006}, and fully self-consistent $GW$  \cite{Ku/Eguiluz:2002,Stan/Dahlen/Leeuwen:2006,Rostgaard/Jacobsen/Thygesen:2010,Caruso/etal:2012,Caruso/etal:2013,Kutepov:2016,Kutepov:2017}.
In addition to the quasiparticle energies, self-consistent schemes provide access to additional quantities such as the charge density, quasiparticle wave functions and the total energy. While some of $GW$'s shortcomings can be remedied by choosing a better starting point \cite{Rinke/etal:2005,Fuchs/etal:2007,Jiang/etal:2009,Marom/etal:2012,Bruneval/Marques:2013} within the ($G_\text{0}W_\text{0}$) scheme, or by going to self-consistent $GW$, there are cases that require a beyond-$GW$ treatment.  For example, in molecules, the energy ordering and energy spacing between different molecular orbitals (MOs)
are erroneously predicted in $GW$, irrespective of the starting point or self-consistent schemes \cite{Marom/etal:2012,Ren/etal:2015}. For open-shell
molecules with pronounced ``static" correlations, a combination of $GW$ and wave-function method in an embeding framework seems to work well
\cite{Dvorak/Rinke:2019,Dvorak/Golze/Rinke:2019}.
In solids, the localized semicore (e.g. $d$) states typically come out too high in energy from $GW$ calculations, and requires going beyond $GW$
to remedy this deficiency \cite{Grueneis/etal:2014}.

\begin{figure*}[!ht]
\begin{picture}(450,90)
\put(0,0){\includegraphics[scale=0.5]{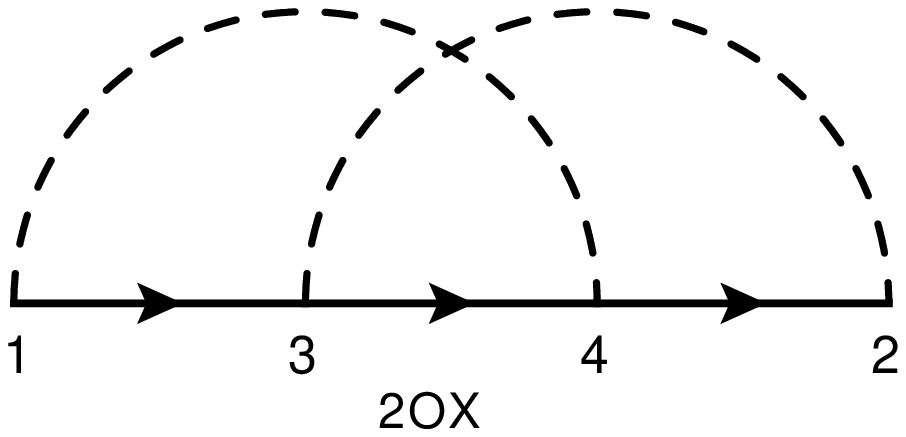}}
\put(5,70){(a)}
\qquad\quad
\put(150,0){\includegraphics[scale=0.5]{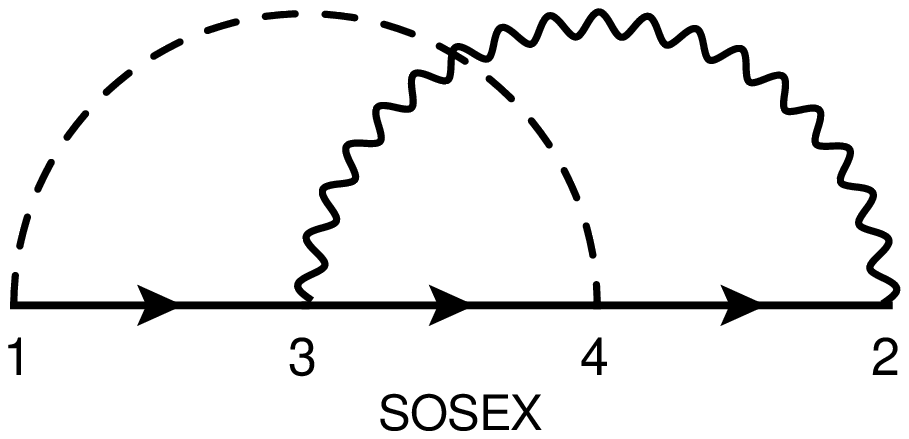}}
\put(155,70){(b)}
\qquad\quad
\put(300, 0){\includegraphics[scale=0.5]{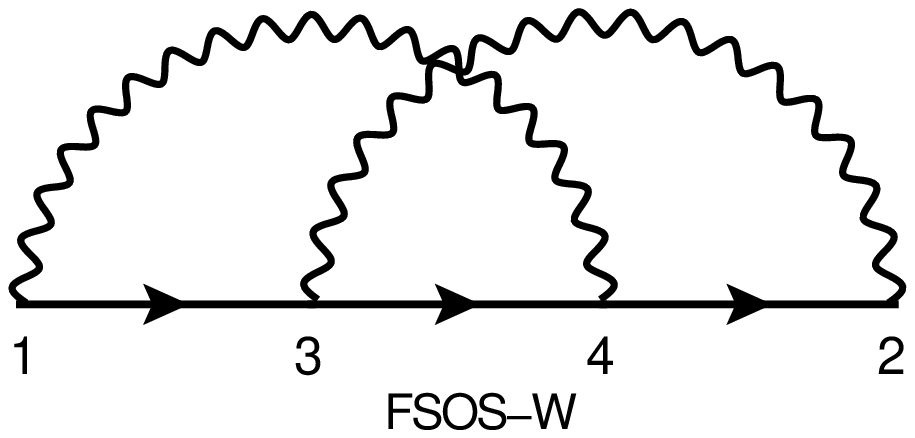}}
\put(305,70){(c)}
\end{picture}
\caption{\label{fig:Feynman-diagrams_expr}
Feynman diagrammatic representations of the 2OX [panel (a)], SOSEX [panel (b)] and FSOS-$W$ [panel (c)] self-energies. Arrowed solid lines, dashed lines, and wavy lines represent the non-interacting Green's function $G_0$, the Coulomb interaction $v$, and the screened Coulomb interaction $W$ respectively.}
\end{figure*}

To overcome the intrinsic shortcomings of $GW$,
one prominent route for further development is to go beyond the $GW$ method by incorporating vertex corrections. ``Vertex corrections" to $GW$ has a long history. Several different types of vertex corrections have been proposed and tested \cite{DelSole/etal:1994,Shirley:1996,Bobbert/Haeringen:1994,Bruneval/etal:2005, Schindlmayr/Godby:1998,Bruneval/etal:2005,Maebashi/Takada:2011,Romaniello/etal:2012,Grueneis/etal:2014,Ren/etal:2015,Kuwahara2016/PhysRevB.94.121116,Kutepov:2016,Kutepov:2017,Maggio/Kresse:2017,Hellgren2018/PhysRevB.98.045117,Vojtech2019/acs.jctc.9b00317,Leeuwen2020/PhysRevB.102.045121}. 
Despite considerable efforts, none of these vertex-corrections have yet found wide-spread use in production calculations. Three factors might be responsible: 
1) Some of the schemes are simply too computationally expensive; 2) the success of certain schemes is limited to particular systems and does not generalize well; 3) the numerical precision of proof-of-principle implementations might not be sufficient so that numerical artifacts mask the actual performance of the method. 

In this article, we present a new formulation and implementation of a second-order screened vertex. Our work is motivated by the $GW$ plus second-order screened-exchange (SOSEX) approach some of the present authors proposed a few years ago \cite{Ren/etal:2015}. In $GW$+SOSEX, the SOSEX self-energy is added to the $GW$ self-energy. The SOSEX self-energy was inspired by the SOSEX correction to the random phase approximation (RPA) for the electron correlation ground state energy \cite{Freeman:1977,Grueneis/etal:2009,Jansen/etal:2010,Paier/etal:2012,Ren/etal:2013}. Diagrammatically, the SOSEX self-energy is similar to the bare second-order exchange (2OX) self-energy \cite{Fetter/Walecka:1971}, albeit  with one bare Coulomb interaction line $v$ replaced by the screened Coulomb interaction line $W$. The diagrammatic representations of the 2OX and SOSEX self-energies are shown in  Fig.~\ref{fig:Feynman-diagrams_expr}(a) and \ref{fig:Feynman-diagrams_expr}(b), respectively. \GnWn+SOSEX outperforms \GnWn for the absolute energy positions of both the highest occupied MO (HOMO) and the lowest unoccupied MO (LUMO), as well as the energy ordering of molecular states. 

While the SOSEX self-energy can be cast into a vertex correction, this vertex cannot be derived from Hedin's equations due to the presence of both the bare and the screened Coulomb interaction in the same diagrams. Instead, the expansion of the electron self-energy in Hedin's equations in powers of $W$ yields at second order diagram that is similar to the direct SOSEX term, but now contains only screened Coulomb interaction lines, as shown in Fig.~\ref{fig:Feynman-diagrams_expr}(c). To distinguish Hedin's second-order self-energy from the SOSEX self-energy given by Fig.~\ref{fig:Feynman-diagrams_expr}(b), we call the former the full second-order self-energy (FSOS) in terms of $W$, abbreviated as FSOS-$W$ in the present work. Adding the FSOS-$W$ correction to the $GW$ self-energy, produces a vertex function $\Gamma$ that is truncated
at first order, i.e., $\Gamma\approx 1+ GGW$, henceforth denoted $\Gamma^{(1)}$. For convenience, we refer to the $GW$+FSOS-$W$ self-energy approximation as \GWG in this work, despite the fact that $\Gamma$ notation for vertex corrections is overused and imprecise. Our \GWG approach, for example, is completely different from what is called {\GnWn}$\Gamma_0$ in Ref.~\cite{Ma/Galli:2019}, where the vertex is given by an exchange-correlation kernel $f_{xc}$ that is included in both $W$ and $\Sigma$.

In the past, the \GWG self-energy scheme has been implemented in the linearized augmented plane-wave (LAPW) framework and applied to simple solids \cite{Kutepov:2016,Kutepov:2017} It was found that vertex corrections in both $W$ and $\Sigma$ are important.  In particular the fully frequency dependent fist-order vertex correction in $\Sigma$ improves bandwidths and band gaps compared to experiment. A static \GWG version was employed by Grueneis \textit{et al.} \cite{Grueneis/etal:2014} to calculate the ionization potentials of solids. However, as will be shown later in this work, approximating the fully dynamic $W(\omega)$ by its static limit $W(\omega=0)$ can significantly underestimate the magnitude of the correction. The behavior of the \GWG approach has also been investigated for the three-dimensional and two-dimensional electron gases, as well as mono-layer and bilayer graphene  \cite{Unimonen/vanLeeuwen:2014,Stefanucci/vanLeeuwen:2015,Pavlyukh:vanLeeuwen:2020}. There it was found that the spectral function is not guaranteed to be positive. A positive definite variant was then proposed \cite{Unimonen/vanLeeuwen:2014,Stefanucci/vanLeeuwen:2015,Pavlyukh:vanLeeuwen:2020}.  
Finally, we note that a correlation energy contribution
corresponding to FSOS-$W$ has been employed as an addition to the RPA correlation energy and applied to  
molecules \cite{Eshuis/Bates/Furche:2012}. Although a self-consistent \GWG scheme can be defined,  
we focus in this work on the one-shot \GnWnG scheme and check its performance for molecular systems, 
based on different starting points.
The obtained results will be compared to those obtained using the \GnWn and \GnWn+SOSEX schemes.


The rest of the paper is organized as follows: in Section~\ref{sec:method} we review the FSOS-$W$ formalism in Hedin's equations and give its concrete expression as implemented in the code, for which the resolution-of-identity technique is employed. The detailed algorithm behind our implementation is described in Appendix.~\ref{appendix:implemn}. In Section~\ref{sec:resl_discuss}, the performance of \GnWnG with different starting points is assessed for selected
molecular test sets and compared to that of \GnWn and \GnWn+SOSEX. The influence of the static approximation is
also discussed. Finally we conclude our work in Section~\ref{sec:conclu-outlook}.

\section{\label{sec:method} Methodology}

\subsection{\label{subsec:theo_bkgd} Theoretical formulation}

Within the Hedin's formulation, the self-energy of an interacting many-electron system is given by
\begin{align}
\Sigma(1,2) &= i\int d3 d4\, G(1,3) W(4,1) \Gamma(3,2,4), \label{eq:selfEnergy_formul}
\end{align}
where numerals denote composite space, spin and time variables, e.g.,  $1=(\mathbf{r}_1,\, \sigma_1,\, t_1)$. In
Eq.~\ref{eq:selfEnergy_formul} the vertex function $\Gamma$ encodes all the many-body complexity that 
goes beyond the $GW$ approximation. Consequently, through Eq.~\ref{eq:selfEnergy_formul}, 
any approximation for $\Gamma$ translates into a corresponding approximation for the self-energy. 

The 3-point vertex function is formally given by 
\begin{eqnarray}
    \Gamma(1,2,3) &=& \delta(1,2)\delta(1,3) + \nonumber \\
                   & &\int d6 d7 \frac{\delta\Sigma(1,2)}{\delta G(4,5)}G(4,6)G(7,5)\Gamma(6,7,3)\, .
    \label{eq:gamma_function_eq}
\end{eqnarray}
Starting with the identity approximation for the vertex, $\Gamma(6,7,3)\approx \delta(1,2)\delta(1,3)$, produces the $GW$ approximation. At the next iteration, we need to evaluate the functional derivative of the $GW$ self-energy \begin{equation}
    \frac{\delta\Sigma(1,2)}{\delta G(4,5)}=iW(1,2)\delta(1,4)\delta(2,5)+G(1,2)\frac{\delta W(1,2)}{\delta G(4,5)}.
\end{equation}
Inserting this functional derivative into the vertex function in Eq.~\ref{eq:gamma_function_eq} gives
\begin{equation}
\Gamma(1,2,3) = \delta(1,2)\delta(1,3) + iW(1,2)G(1,3)G(3,2) + \ldots
\end{equation}
where we omitted terms of order $O(W^2)$ and higher. At first order in $W$ the vertex is thus
\begin{equation}
\Gamma^{(1)}(1,2,3) 
              = \delta(1,2)\delta(1,3) + iW(1,2)G(1,3)G(3,2). \qquad\label{eq:vertex_formul}\,
\end{equation}
and inserting it into the self-energy expression in  Eq.~\ref{eq:selfEnergy_formul} we obtain the \GWG self-energy
\begin{equation}
\Sigma^{\text{\scriptsize \GWG}}(1,2) = \Sigma^{GW}(1,2) + \Sigma^{\text{\scriptsize FSOS-}W}(1,2) \, .
\label{eq:selfe_GWG1}
\end{equation}
$\Sigma^{GW}$ is the $GW$ self-energy, and
\begin{equation}
\Sigma^{\text{\scriptsize FSOS-}W}(1,2) = -\int d3 d4\, G(1,3)W(4,1) W(3,2)G(3,4)G(4,2) 
\label{eq:FSOS_formul}
\end{equation}
is the FSOS-$W$ self-energy. It should be emphasized that a perturbative expansion of the self-energy in terms of $W$ produces only 
one diagram at first order ($GW$), and one at the 2nd order (FSOS-$W$), but six at the 3nd order. The self-energy diagrams up to the
first three orders were already been
presented in the seminal $GW$ paper by Hedin in 1965 \cite{Hedin:1965}, and included in Appendix~\ref{appendix:diagram} 
of the present work for completeness.

Equation~\ref{eq:FSOS_formul} is \textit{de facto} the mathematical expression behind the diagrammatic 
representation of Fig.~\ref{fig:Feynman-diagrams_expr}(c). In contrast, as discussed in Ref.~\cite{Ren/etal:2015}, 
the mathematical expression for the SOSEX self-energy, corresponding to Fig.~\ref{fig:Feynman-diagrams_expr}(b), is given by
\begin{equation}
\Sigma^{\text{\scriptsize SOSEX}}(1,2) =  
 -\int\text{d}3\text{d}4\, G(1,3)v(4,1) W(3,2)G(3,4)G(4,2). \label{eq:sosex_formul}
\end{equation}

As mentioned in the introduction, we are only interested in one-shot \GnWnG, i.e., in using only the  noninteracting reference Green's function $G_0$ and the corresponding screened Coulomb interaction $W_0$ without iterating eq.~\ref{eq:FSOS_formul} to self-consistency via Dyson's equation. The FSOS-$W$ self-energy of Eq.~\ref{eq:FSOS_formul} calculated using $G_0$ and $W_0$ will be denoted FSOS-$W_0$.

Starting from the reference Kohn-Sham density functional theory (KS-DFT) or Hartree-Fock (HF) orbitals $\psi_n(\mathbf{r})$ 
and orbital energies $\varepsilon_n$, the non-interacting Green's function on the imaginary frequency axis is given by
\begin{equation}
G_0(\mathbf{r}, \mathbf{r}'; i\omega) = \sum_n \frac{ \psi_n(\mathbf{r}) \psi^*_n(\mathbf{r}') }
{i\omega + \varepsilon_{\text{F}} -\varepsilon_n}, \label{eq:Green-func_expr}
\end{equation}
where $\varepsilon_{\text{F}}$ is the Fermi level. The screened Coulomb interaction $W_0$ can be conveniently represented
within the products of the single-particle molecular orbitals (MOs), resulting in the two-electron screened Coulomb repulsion integrals,
\begin{eqnarray}
&&\langle pq| W_0(i\omega) |rs \rangle \nonumber \\
&&\enspace= \int \text{d}\mathbf{r} \text{d}\mathbf{r}'\, 
\psi^{*}_p(\mathbf{r}) \psi_r(\mathbf{r}) W_0(\mathbf{r}, \mathbf{r}'; i\omega) 
\psi^*_q(\mathbf{r}') \psi_s(\mathbf{r}'). \label{eq:4centr-intgr_formul}
\end{eqnarray}
where $p,q,r,s$ denote arbitrary (both occupied and virtual) MOs.
Using Eqs.~\ref{eq:FSOS_formul}, \ref{eq:Green-func_expr} and \ref{eq:4centr-intgr_formul}, one
can readily obtain the diagonal elements of the FSOS-$W_0$ self-energy along the imaginary frequency axis,
\begin{align}
&\Sigma^{\text{\scriptsize FSOS-}W_0}_n(i\omega) = \int d\bfr d\bfrp \psi_n(\mathbf{r}) 
\Sigma^{\text{\scriptsize FSOS-}W_0}(\mathbf{r}, \mathbf{r'}; i\omega) \psi_n(\mathbf{r}'), 
\nonumber \\
&=\sum_{p,q,r} \int \frac{\text{d}\omega'}{2\pi} 
\frac{(f_p - f_q)\langle nq| W_0(i\omega') |rp\rangle \langle rp| W_0(i\omega + i\omega') 
|qn \rangle}
{ (i\omega + i\omega' + \varepsilon_{\text{F}} - \varepsilon_r)(i\omega' + \varepsilon_p 
- \varepsilon_q) }, \label{eq:FSOS_expr}
\end{align}
where $f_p$ and $f_q$ are the occupation numbers of the states $p$ and $q$, respectively.

\subsection{\label{subsec:implementation}Implementation}

The computational expression of the FSOS-$W_0$ self-energy given in Eq.~\ref{eq:FSOS_expr} is implemented in the all-electron Fritz-Haber-Institute \textit{ab initio} simulations (FHI-aims) package~\cite{Blum2009,Havu/etal:2009,Ren/etal:2012}. In FHI-aims, both numeric atom-centered orbitals (NAOs) and gaussian-type orbitals (GTOs) can be used as basis functions. The single-particle MOs $\lbrace \psi_p(\mathbf{r}) \rbrace$ are expanded in terms of a set of atomic orbital (AO) basis functions $\lbrace \phi_j(\mathbf{r})\rbrace$,
\begin{equation}
\psi_p(\mathbf{r}) = \sum_j c_{pj} \phi_j(\mathbf{r})\, . \label{eq:single-particle_exp} \\
\end{equation}
Similar to the $GW$ and SOSEX implementation in FHI-aims \cite{Ren/etal:2012,Ren/etal:2015}, the resolution-of-identity (RI) approximation with the Coulomb metric
\cite{Dunlap/Connolly/Sabin:1979,Whitten:1973,Vahtras/Almlof/Feyereisen:1993} is used to compute the two-electron screened Coulomb integrals given in
Eq.~\ref{eq:4centr-intgr_formul}. With the RI approximation, a set of auxiliary basis functions (ABFs) $\{ P_{\mu}(\mathbf{r}) \}$ is introduced 
to represent the pair products of AO basis functions,
\begin{equation}
\phi_i(\mathbf{r}) \phi_j(\mathbf{r}) \approx \sum_{\mu} C^{\mu}_{ij} P_{\mu}(\mathbf{r}), 
\label{eq:product-basis_expr}
\end{equation}
where the expansion coefficients $\{C^{\mu}_{ij}\}$ are determined using the Coulomb metric \cite{Dunlap/Connolly/Sabin:1979,Whitten:1973,Vahtras/Almlof/Feyereisen:1993,Ren/etal:2012} in the present implementation. 
Here and in the following, the Greek letters $\mu,\nu,\alpha,\beta$ denote ABFs.
To evaluate the screened Coulomb integrals, we
need to further transform the 3-index expansion coefficients from the AO basis to MO orbitals, namely,
\begin{equation}
M^{\mu}_{pr} = \sum_{i,j} c^*_{pi}c_{rj} C^{\mu}_{ij}\, .
\label{eq:triple_MO}
\end{equation}
Now, using Eqs~\ref{eq:4centr-intgr_formul} and \ref{eq:single-particle_exp}-\ref{eq:triple_MO}, we obtain
\begin{equation}
\langle pq| W_0(i\omega) |rs\rangle =\sum_{\mu,\nu} M^{\mu}_{pr} W_{0,\mu\nu}(i\omega) 
M^{\nu}_{qs}\, , \label{eq:scr_2eri_RI}
\end{equation}
where 
\begin{equation}
    W_{0,\mu\nu}(i\omega) = \int d\bfr d\bfrp P_\mu(\bfr) 
    W_0(\bfr, \bfrp, i\omega) P_\nu(\bfrp)\, .
    \label{eq:W_0_matr_def}
\end{equation}

To obtain the matrix form of $W_0$ within the set of ABFs, we invoke the geometrical series expansion of
$W_0$ in terms of $v$ and the non-interacting response function $\chi_0(i\omega)$, i.e., $W_0=V+V\chi_0(i\omega) V + \cdots$.
Introducing $\Pi(i\omega) = V^{1/2} \chi_0(i\omega) V^{1/2}$, it is straightforward to see that
\begin{equation}
    W_0(i\omega) = V^{1/2} \left[1-\Pi(i\omega) \right]^{-1} V^{1/2}
    \label{eq:W_0_operator}
\end{equation}
where $V$, $\chi_0$ and $\Pi$ here, similar to $W_0$, are matrices within the auxiliary basis set. Here
\begin{equation}
V_{\mu\nu} = \int \text{d}\mathbf{r} \text{d}\mathbf{r}'\, \frac{ P_{\mu}(\mathbf{r}) 
P_{\nu}(\mathbf{r}') }{|\mathbf{r} - \mathbf{r}'|}, \label{eq:Coulomb_aux-expr}
\end{equation}
is the Coulomb matrix, and $V^{1/2}$ is the square root of the $V$ matrix. 
To derive the computational expression for the $\Pi$ matrix, one should be aware that the $\chi_0$ matrix
is defined to represent the real-space expression of $\chi_0$,
\begin{equation}
    \chi_0(\bfr,\bfrp,i\omega) = \sum_{\mu,\nu}P_\mu(\bfr) \chi_{0,\mu\nu}(i\omega) P_\nu(\bfrp)\, .
\end{equation}
which is given by,
\begin{equation}
    \chi_0(\bfr,\bfrp,i\omega) = \sum_{m,n}
    \frac{(f_m-f_n)\psi^\ast_m(\bfr)\psi_n(\bfr)\psi_n^\ast(\bfrp)\psi_m^\ast(\bfrp)}{\epsilon_m-\epsilon_n-i\omega} \, ,
\end{equation}
Using Eqs.~\ref{eq:product-basis_expr} and \ref{eq:triple_MO}, one immediately arrives at,
\begin{equation}
       \chi_{0,\mu\nu}(i\omega) = \sum_{m,n}
    \frac{(f_m-f_n)M_{mn}^{\mu}M_{nm}^{\nu}}{\epsilon_m-\epsilon_n-i\omega} \, .
\end{equation}

For convenience, one may further introduce another set of 3-index integrals by
multiplying $V^{1/2}$ with $M$,
\begin{equation}
O^{\mu}_{pr} = \sum_{i,j}\sum_{\nu} V^{1/2}_{\mu\nu} M^{\nu}_{pr} \, , 
\label{eq:3integro_aux-expr}
\end{equation}
and it follows that
\begin{equation}
           \Pi_{\mu\nu}(i\omega) = \sum_{m,n}
    \frac{(f_m-f_n)O_{mn}^{\mu}O_{nm}^{\nu}}{\epsilon_m-\epsilon_n-i\omega} \, .
\end{equation}
Using Eqs.~\ref{eq:scr_2eri_RI}, \ref{eq:W_0_operator} and \ref{eq:3integro_aux-expr}, we eventually obtain
\begin{equation}
\langle pq| W_0(i\omega) |rs\rangle = \sum_{\mu,\nu} O^{\mu}_{pr} [\, 1- \Pi(i\omega) \,]^{-1}_{\mu\nu} 
O^{\nu}_{qs}\, . \label{eq:4centr-intgr_expr}
\end{equation}
Finally, based on Eq.~\ref{eq:FSOS_expr} and \ref{eq:4centr-intgr_expr}, 
the FSOS-$W_0$ self-energy is expressed as,
\begin{eqnarray}
&&\Sigma^{\text{\scriptsize FSOS-}W_0 }_n (i\omega) = \sum_{p,q,r} \sum_{\mu,\nu} \sum_{\alpha,\beta} (f_p - f_q) 
\int^{\infty}_{-\infty} \frac{ \text{d}\omega' }{ 2\pi } \nonumber \\
&&\frac{ O^{\mu}_{nr} [\, 1- \Pi(i\omega') \,]^{-1}_{\mu\nu} O^{\nu}_{qp} 
O^{\alpha}_{rq} [\, 1- \Pi(i\omega + i\omega') \,]^{-1}_{\alpha\beta} O^{\beta}_{pn} }
{ (i\omega + i\omega' + \varepsilon_{\text{F}} -\varepsilon_r)(i\omega' + \varepsilon_p 
- \varepsilon_q) }\, . \quad\enspace \label{eq:FSOS_implemn}
\end{eqnarray}

The detailed algorithm for implementing Eq.~\ref{eq:FSOS_implemn}, including the loop structure and
the parallelization strategy, is presented in Appendix~\ref{appendix:implemn}. The canonical scaling
of a straightforward implementation of Eq.~\ref{eq:FSOS_implemn} is $O(N^5)$ with $N$ being the
system size, but practically our 
implementation achieves a $O(N^{4.1})$ scaling for the range of system sizes we have looked at, as demonstrated in
Ref.~\ref{subsec:scal_performance}. Detailed benchmark
results and discussions will be presented in Sec. \ref{subsec:scal_performance}.


In our present implementation, the \GnWnG self-energy 
is first evaluated on the imaginary frequency axis, and
then analytically continued to the real frequency axis. To this end, the Pad$\acute{\text{e}}$ approximant
is used, where the self-energy is given by Thiele's reciprocal difference method~\cite{Baker1975},
\begin{eqnarray}
\Sigma^{\GnWnG}(i\omega) &&= \cfrac{ a_1 }{ 1+ \cfrac{ a_2(i\omega - i\omega_1) }{ 1+ \cfrac{ a_3(i\omega - i\omega_2) }{ 1+ \cdots } } }
\end{eqnarray}
with $\{ i\omega_k, \Sigma^{\GnWnG}(i\omega_k),\, k=1, \cdots, n \}$ being the chosen interpolation points, obtained from practical $\GnWnG$
calculations. The coefficients of the interpolant $\{a_k\}$ are determined by the recursive relation.
The analytic continuation for each of the diagonal self-energy element is done by substituting $i\omega$ with $\omega$ directly, namely,
\begin{eqnarray}
&&\Sigma^{\text{\scriptsize \GnWnG}}(i\omega)
 \xrightarrow{ \text{analy. continu.} } \enspace\Sigma^{\text{\scriptsize \GnWnG}}(\omega).
\end{eqnarray}
Finally, the \GnWnG quasiparticle (QP) energy can be determined iteratively, 
\begin{eqnarray}
&&\epsilon^{\text{\scriptsize QP}}_n = \epsilon_n + 
  \langle n| \text{Re}[\Sigma(\epsilon^{\text{\scriptsize QP}}_n )] -V_{\text{\scriptsize xc}} |n \rangle.
 \label{eq:qp_eqn}
\end{eqnarray}

\subsection{\label{subsec:scal_performance}Scaling behavior}

\begin{figure*}[!ht]
\includegraphics[scale=1]{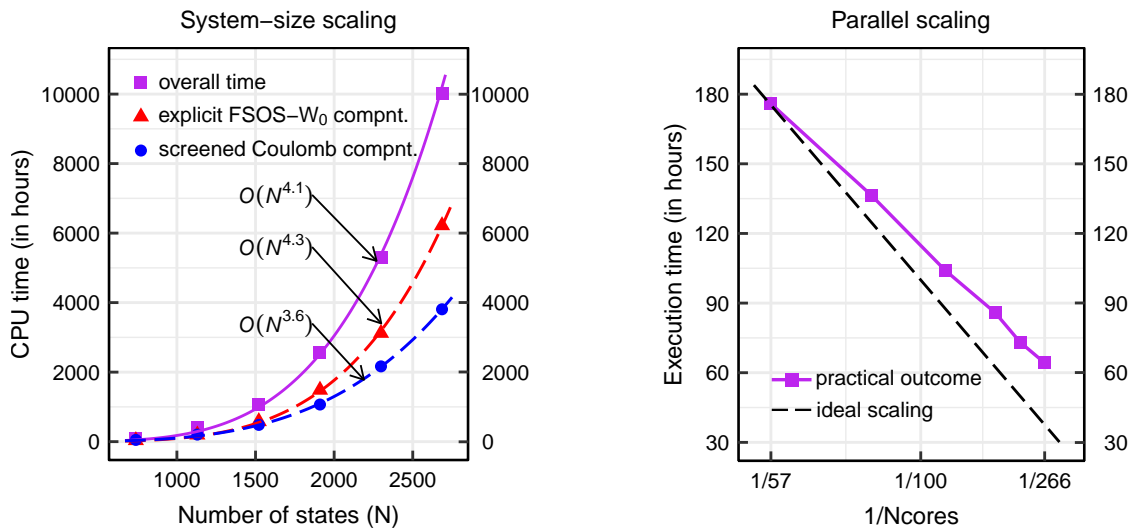}
\caption{\label{fig:scal_perform}
The scaling behavior for the computational time of  FSOS-$W_0$ self-energy calculations with regard to system size (left panel) and the number of CPU cores (right panel). The test systems are linear oligoacenes C$_{4n+2}$H$_{4n+2}$ with the six data points on the left panel corresponding to
$n=1$ to 6. The timing data are fitted by power-law functions $y(x)=ax^b$ with $x$ being the total number of states of the system ($N$). The
(blue) circles denote the time cost for evaluating the $\Pi$ matrix and its inverse $(1- \Pi)^{-1}$ (referred to as the ``screened Coulomb component" in the plot). The (red) triangles refer to the time cost for evaluating Eq.~\ref{eq:FSOS_implemn} (labelled as ``explicit FSOS-$W_0$ component"). 
The aug-cc-pVQZ basis set and 60 frequency points are used in the calculations, and the QP energies of all occupied states are 
computed. All calculations are done on machines with Intel(R) Xeon(R) Platinum 9242 CPU $@$2.30GHz cores. }
\end{figure*}
A straightforward implementation of Eq.~\ref{eq:FSOS_implemn} is computationally demanding, since the dynamically screened Coulomb interaction needs to be evaluated at two different frequency points. By carefully designing the loop structure and manipulating the different terms (see Appendix~\ref{appendix:implemn}), our implementation makes \GnWnG calculations affordable and facilitates the application of \GnWnG to interesting molecular systems with tens of atoms. 

Here we demonstrate the scaling behavior of the computational cost (in terms of CPU hours) with respect to the system size, as well as the parallel efficiency with respect to the number of CPU cores. For this purpose, we choose the oligoacenes 
(C$_{4n+2}$H$_{4n+2}$) from benzene ($n=1$) to hexacene ($n=6$) as test systems. The gaussian aug-cc-pVQZ 
\cite{Dunning:1989} basis set is used, leading to 2686 one-electron basis functions for $n=6$. 

In the left panel of Fig.~\ref{fig:scal_perform}, the CPU time as a function of the total number of states $N$ is shown. The three data sets correspond to the total timing, the timing for evaluating Eq.~\ref{eq:FSOS_implemn}, and that for calculating the screened Coulomb matrix, respectively. The solid or dashed lines are power-law fits. The number of molecular states $N$ shown on the $x$-axis is the sum of occupied ($N_\text{occ}$) and unoccupied ($N_\text{unocc}$) states, and, in our calculations, equals the number of one-electron basis functions.


Figure~\ref{fig:scal_perform} reveals that the overall scaling with system size is $O(N^{4.1})$ and thus better than the formal scaling of O($N^5$). The formal scaling of the \GnWnG method can be inferred from the rate-determining step in the evaluation of the FSOS-$W_0$ part of the self-energy in Eq.~\ref{eq:FSOS_implemn}. Specifically, as shown in the computational algorithm schematic in Fig.~\ref{fig:fsos_algom} in Appendix~\ref{appendix:implemn}, this step scales as 
$O(N_n N N_{\text{occ}} N_{\text{unocc}} N_{\text{aux}})$ where  $N_\text{occ}$ and $N_\text{unocc}$ are the numbers of occupied and unoccupied states, $N_{\text{aux}}$ is the number of ABFs and $N_n$ the number of quasiparticle states to be calculated. Since $N_n$, $N_\text{occ}$, $N_{\text{unocc}}$, and $N_{\text{aux}}$ are all proportional to $N$, the formal scaling should be $O(N^5)$.

The $O(N^{4.1})$ scaling of our implementation arises from the constructing of $\Pi$ matrix and the screened Coulomb interaction $W_0$ (i.e., the matrix inversion operation) in the large auxiliary basis space involving two frequency variables. These steps are computationally heavy but with a lower scaling ( $O(N^{4.0})$ for $\Pi$ and $O(N^{2.9})$ for $(1-\Pi)^{-1}$). Due to the carefully designed loop structure, the final step of assembling the  FSOS-$W_0$ self-energy scales as $O(N^{4.3})$ for the tested system sizes, as indicated by the red dashed line in the left panel of Fig.~\ref{fig:scal_perform}. 

We also parallelized our \GnWnG implementation. To check the parallel efficiency, we take the hexacene molecule as a test system, and present in the right panel of Fig.~\ref{fig:scal_perform} the execution time as a function of the number of CPU cores employed in the calculations.  
As can be seen from the plot, our parallel implementation scales up to several hundreds of CPU cores. From 57 to 266 CPU cores, the wall-clock time is
reduced by a factor of 2.3, indicating a parallel efficiency of approximately $50\%$ with the number of cores increasing by 5 times. 
In brief, our implementation renders it feasible to perform \GnWnG calculations for molecules containing a few tens of atoms 
with moderate computational resources.

\section{\label{sec:resl_discuss}Results and Discussions}

In this section, we examine the performance of \GnWnG for quasiparticle energies, in particular the
HOMO and LUMO levels for selected molecular test sets. For this purpose, we take two molecular test sets: 
the $GW$100 set \cite{Setten/etal:2015} and  
a set of 24 organic acceptor molecules (termed  Acceptor24 in the following) \cite{Richard2016/10.1021/acs.jctc.5b00875,Knight/etal:2016}. While $GW100$ contains a diverse set of closed shell atoms and small molecules, Acceptor24 provides medium-sized molecules. In addition, we investigate the energy ordering of the benzene molecule and screening effects in the CO molecule. The reference results used here to benchmark \GnWnG are taken from the literature, obtained either with the coupled-cluster method with singles, doubles, and perturbative triples [CCSD(T)] excitations \cite{Katharina/Klopper:2015,Richard2016/10.1021/acs.jctc.5b00875} or the equation-of-motion coupled-cluster with single and double excitations [EOM-CCSD] \cite{Lange/Berkelbach:2018}.

Similar to \GnWn, the \GnWnG method also depends on the reference state on which it is based. To obtain a more complete picture of the behavior of \GnWnG, in this work we shall carry out benchmark \GnWnG calculations with three different starting points, namely, PBE \cite{Perdew/Burke/Ernzerhof:1996}, PBE0 \cite{Perdew/Ernzerhof/Burke:1996,Carlo1999} and HF. 

For $GW100$, the Gaussian basis set def2-TZVPP \cite{Florian2005} is employed, to be consistent with the coupled-cluster calculations from which the reference results are obtained \cite{Rangel2016,Lange/Berkelbach:2018}. For the Acceptor24 set, the FHI-aims-2009 \cite{Blum2009} \textit{tier} 4 basis set is used, to be consistent with the choice made in Ref.~\cite{Knight/etal:2016} where the Acceptor24 test set was originally proposed. Previous benchmark calculations show that  \textit{tier} 4 comes to within 0.1 eV of
the complete basis set limit \cite{Ren/etal:2012,Ren/etal:2015}.

As mentioned above, the Pad$\acute{\text{e}}$ approximant~\cite{Gunnarsson2010,Blase2020} is for analytically continuing the self-energy from the imaginary frequency axis to the real one. Instead of fixing the number of Pad$\acute{\text{e}}$ parameters empirically, we increase the number of Pad$\acute{\text{e}}$ parameters as well as the number of frequency grid points to ensure that the quasiparticle energies are converged. We had to exclude a couple of systems from the reported results, because we could not obtain stable QP energies from the Pad$\acute{\text{e}}$-based analytical continuation.

\subsection{Ionization potentials and electron affinities of the \textit{GW}100 test set}

$GW$100 is \cite{Setten/etal:2015} a test set that contains 100 small closed-shell atoms and molecules, first assembled by {van Setten} \textit{et al.}\cite{Setten/etal:2015}. $GW$100 provides accurate {\GnWn} vertical ionization potentials (IPs) and 
electron affinities (EAs) computed with three different $GW$ codes: FHI-aims, Turbomole \cite{Setten/etal:2013}, and BerkeleyGW \cite{Deslippe/etal:2012}. In the following, IP and EA refer to \textit{vertical} IPs and EAs and not  \textit{adiabatic} ones.] These were followed by benchmark calculations using VASP \cite{Maggio/Kresse:2017} and WEST \cite{Ma/Galli/etal:2019}, and a few other codes that have $GW$ implementations like ABINIT, CP2K, Fiesta, MOLGW and YAMBO
\cite{GW100press}. 
$GW$100 is thus a natural choice to benchmark our \GnWnG scheme. 

In the present work, we chose to benchmark \GnWnG against quantum chemistry methods (CCSD(T) or EOM-CCSD) and not against experiment. In addition, we fix the one-electron basis and do not extrapolate the results to the complete basis set (CBS) limit. This all-theory comparison has the advantage that it is internally consistent. To then compare our results to experiment would require CBS extrapolation and the addition of vibrational and zero-point motion effects.

\begin{figure*}[!ht]
\includegraphics[scale=1]{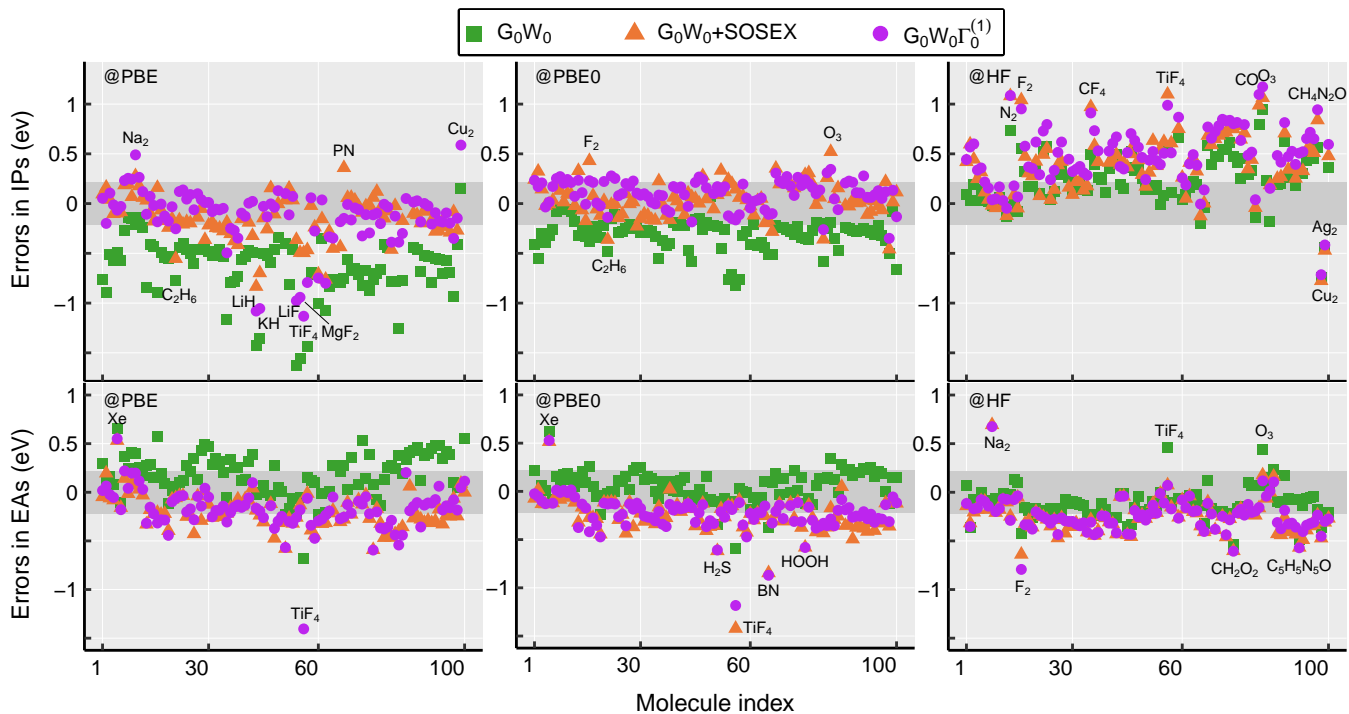}
\caption{\label{fig:GW100_IPs_EAs_accuracy}
Deviation of the $G_0W_0$, $G_0W_0$+SOSEX, and \GnWnG IPs (upper panels) and EAs (lower panels) of the $GW$100 test from their respective CCSD(T) \cite{Katharina/Klopper:2015} [for IPs] or EA-EOM-CCSD \cite{Lange/Berkelbach:2018} [for EAs)] reference values. Results for three different starting points -- PBE, PBE0, and HF -- are presented in the left, middle, and right panels, respectively.
The dark grey shaded area indicates an error range of $[-0.2, 0.2]$ eV. The outlier molecules with large errors are labelled by their chemical
formulas.
}
\end{figure*}

In Fig.~\ref{fig:GW100_IPs_EAs_accuracy}, we present the error distributions of the $GW$100 IPs (upper panels) and EAs (lower panels) computed with
\GnWn, \GnWn+SOSEX, and \GnWnG. The IPs are referenced to CCSD(T) \cite{Katharina/Klopper:2015,Richard2016/10.1021/acs.jctc.5b00875} and the EAs to EOM-CCSD \cite{Lange/Berkelbach:2018}. For each method, we used three different starting points and the corresponding results (error distributions) are presented in the left (PBE), middle (PBE0), and right (HF) panels of Fig.~\ref{fig:GW100_IPs_EAs_accuracy}. The IP and EA values are tabulated in Tables~S1 and S2 of the Supporting Information (SI). The difference between our \GnWn@PBE values and those reported in Ref.~\cite{Setten/etal:2015} are mainly due to the different basis sets used. One should note that the errors presented in Fig.~\ref{fig:GW100_IPs_EAs_accuracy} are obtained by subtracting the reference CCSD(T)/EOM-CCSD values from the \GnWn or beyond-\GnWn values. Hence a negative IP or EA error indicates that the computed $GW$-type  IP or EA value is too small, corresponding to a too-high QP HOMO or LUMO level, and vice versa. The mean signed deviations (MSDs) and the mean absolute deviations (MADs) of various computational methods for 
the entire $GW$100 set are presented in Table~\ref{tab:GW100_IPsEAs_MDs}. For completeness, we also present in Table~S3 of the Supporting Information the MSDs and MADs for the IPs with the EOM-CCSD results as the reference. We note that the MSD an MAD values in Table~S3  do not differ much from those presented in Table~\ref{tab:GW100_IPsEAs_MDs}, indicating that CCSD(T) and EOM-CCSD reference results are very close.

We first examine the @PBE results, since PBE is the most popular starting point for \GnWn and beyond calculations. We observe that \GnWn@PBE generally underestimates the IPs and overestimates the EAs, meaning that QP HOMOs are too high and QP LUMOs too low. This behavior is characteristic of \GnWn@PBE, and has been observed and discussed before in the literature \cite{Ren/etal:2012,Setten/etal:2013,Ren/etal:2015,Setten/etal:2015,Marom:2017}. Upon adding vertex corrections, the IP significantly improve, and the MSD (MAD) is reduced from $-0.66$ (0.66) eV to $-0.15$ (0.20) eV for (\GnWn+SOSEX)@PBE and to $-0.13$ (0.21) for \GnWnG@PBE. Inspection of Fig.~\ref{fig:GW100_IPs_EAs_accuracy} reveals that the too high QP HOMOs are pushed down roughly to the right energy range. Interestingly, the overall error statistics is almost the same for the two vertex-corrected beyond-\GnWn schemes, although individual IP and EA values differ slightly.  For EAs, \GnWn@PBE only overshoots slightly, giving a MSD (MAD) of 0.19 (0.24) eV. The SOSEX or FSOS-$W$ corrections push the QP LUMOs up, yielding an overall underestimation of the EAs, with a negative MSD of $-0.21$ eV for (\GnWn+SOSEX)@PBE and $-0.17$ eV for \GnWnG@PBE. The MADs for all three methods, with the PBE starting point, are very close, ranging from 0.21 eV to 0.24 eV. 


Inspecting @PBE0 next, we see that \GnWn@PBE0 improves significantly on \GnWn@PBE. The IP MSD and MAD are drastically reduced by a factor of 2. Also \GnWn+SOSEX and \GnWnG improve the IPs, but only slightly. Now the majority of data points (i.e., the deviations from the reference values) lie within the desired energy window
of $[-0.2, 0.2]$ eV, as indicated by the shaded area in Fig.~\ref{fig:GW100_IPs_EAs_accuracy}. This observation is also corroborated by the reduced MSD and MAD values shown in Table~\ref{tab:GW100_IPsEAs_MDs}. For the EAs, \GnWn@PBE0 again improves noticeably over \GnWn@PBE, as the too low LUMO levels are pushed up. However, the \GnWn+SOSEX and \GnWnG EA values do not show an appreciable change.

Finally, for the HF reference state, the evolution of the \GnWn results continues. HOMO levels move downwards and the LUMO levels upwards. In the end, \GnWn@HF overestimates IPs as much as \GnWn@PBE0 underestimates them, whereas the situation for EAs is reversed. On the contrary, IPs are clearly overestimated by \GnWn+SOSEX and \GnWnG based on HF. EAs, however, do not change substantially, compared to those obtained with PBE or PBE0 starting points. In brief, although HF appears to be a reasonable choice for \GnWn for small closed-shell molecules such as those included in $GW$100, this is clearly not the case for \GnWn+SOSEX and \GnWnG.

After discussing the general behaviour of $G_0W_0$, $G_0W_0$+SOSEX, and \GnWnG for the $GW$100 test set, we now briefly discuss some of the outliers visible in  Fig.~\ref{fig:GW100_IPs_EAs_accuracy}. For example, for TiF$_4$ and CH$_4$O \GnWnG@PBE underestimates the EAs by $1.4$ and $1.3$ eV, respectively. For TiF$_4$, we further observe that the \GnWn and beyond-\GnWn results are highly sensitive to the starting point, probably due to the presence of the transition metal atom Ti. Going from the PBE to HF reference state, the calculated \GnWnG EA value is significantly improved. This is in contrast with most other cases where the \GnWnG EA value does not change much by varying the reference state. 
As indicated in Fig.~\ref{fig:GW100_IPs_EAs_accuracy}, there are a few more outliers for which  \GnWn+SOSEX or \GnWnG results still show sizable
deviation from the reference values. The origin for the failure of the vertex corrections for these ``problematic cases" needs to be further
investigated in future work.

\begin{table*}[!ht] 
\caption{\label{tab:GW100_IPsEAs_MDs}MSDs (left sub-columns) and MADs 
(right sub-columns) of the $GW100$ set with respect to the IP-CCSD(T)~\cite{Katharina/Klopper:2015} and EA-EOM-CCSD~\cite{Lange/Berkelbach:2018} results, respectively, for vertical IPs and EAs computed with the def2-TZVPP basis. ``SP" is the abbreviation for ``starting point".}
\begin{ruledtabular}
\begin{tabular}{l dd dd dd d dd dd dd}
&\multicolumn{6}{c}{IPs} & &\multicolumn{6}{c}{EAs} \\
\colrule
\multicolumn{1}{l}{\diagbox[width=1.9cm, height=0.78cm]{\raisebox{-0.3\height}{SP}}{Method}}
&\multicolumn{2}{c}{\textrm{$G_0W_0$}}     &\multicolumn{2}{c}{\textrm{$G_0W_0$+SOSEX}}     
&\multicolumn{2}{c}{\textrm{$G_0W_0\Gamma^{(1)}_0$}}
&
&\multicolumn{2}{c}{\textrm{$G_0W_0$}}     &\multicolumn{2}{c}{\textrm{$G_0W_0$+SOSEX}}     
&\multicolumn{2}{c}{\textrm{$G_0W_0\Gamma^{(1)}_0$}}   \\
\colrule
@PBE     &-0.66 &0.66    &-0.15 &0.20    &-0.13 &0.21    &    & 0.19 &0.24    &-0.21 &0.24    &-0.17 &0.21 \\
@PBE0    &-0.31 &0.31    & 0.04 &0.13    & 0.10 &0.15    &    & 0.06 &0.13    &-0.25 &0.27    &-0.23 &0.24 \\
@HF      & 0.25 &0.30    & 0.40 &0.43    & 0.49 &0.51    &    &-0.11 &0.16    &-0.25 &0.27    &-0.24 &0.26 \\
\end{tabular}
\end{ruledtabular}
\end{table*}

\subsection{The IPs and EAs for the Acceptor24 test set}

The $GW$100 test set discussed in the above subsection only contains atoms and small molecules. Here we further examine the performance of \GnWnG for Acceptor24 -- a set of 24 molecules of medium size (up to a few tens of atoms) that are relevant in organic electronics.  The Acceptor24 test set was first used to benchmark long-range corrected hybrid functionals \cite{Gallandi/etal:2016}, $GW$-based \cite{Knight/etal:2016}, and electron propagator methods \cite{Dolgounitcheva/etal:2016} against CCSD(T) reference data \cite{Richard2016/10.1021/acs.jctc.5b00875}.
For the $GW$-based methods, (\GnWn+SOSEX)@PBE clearly stood out and outperformed \GnWn based on different starting points as well as fully self-consistent $GW$ for the EAs.

\begin{figure*}[!ht] 
\includegraphics[scale=1]{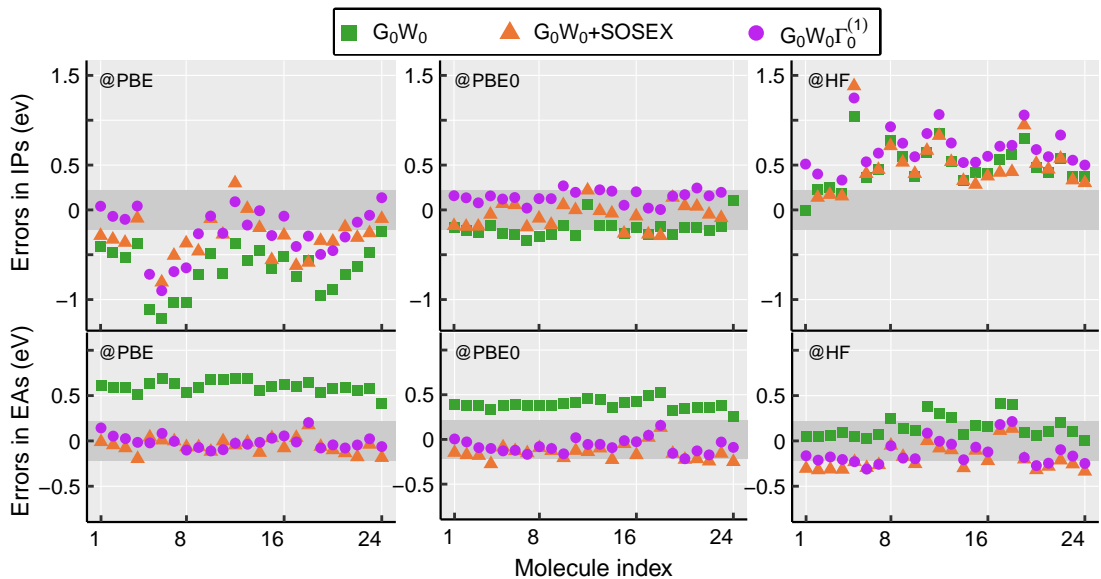}
\caption{\label{fig:Acceptor24_IPs_EAs_accuracy}
Deviation of the \GnWn, \GnWn+SOSEX, and \GnWnG IPs (upper panels) and EAs (lower panels) of the Acceptor24 test set  with respect to CCSD(T) reference data \cite{Richard2016/10.1021/acs.jctc.5b00875}. For each method, results for PBE, PBE0, and HF starting points are presented in the left, middle, and right panels.}
\end{figure*}

In this work, we add \GnWnG to the benchmark. We use the same molecular geometries and basis sets as in Ref.~\cite{Knight/etal:2016}. The \GnWnG IP and EA values based on PBE, PBE0, and HF reference states are presented in Tables~S4 and S5 of the Supporting Information. The deviations from the CCSD(T) reference data for individual molecules are graphically illustrated in Fig.~\ref{fig:Acceptor24_IPs_EAs_accuracy} and the error statistics is presented in Table~\ref{tab:Acceptor24_IPsEAs_MDs}. In analogy to the $GW$100 set, \GnWn and \GnWn+SOSEX results are included for comparison. Small differences in our \GnWn and \GnWn+SOSEX results from those reported in Ref.~\cite{Knight/etal:2016} stem from tighter convergence criteria for the number of frequency points and the number of Pad$\acute{\text{e}}$ parameters in the analytical continuation. These small difference do not affect the assessment of the computational methods studied here.

The overall behaviour of the Acceptor24 IPs is similar to $GW$100 for \GnWn, \GnWn+SOSEX, and \GnWnG. Namely, adding vertex corrections, \GnWn+SOSEX and \GnWnG based on both PBE and PBE0 clearly improve over the corresponding \GnWn schemes. However, both \GnWn+SOSEX and \GnWnG worsen significantly for the HF starting point. \GnWnG is slightly better than \GnWn+SOSEX for the PBE starting point, whereas the opposite when both are based on PBE0. 

For Acceptor24 EAs,  the performance of (\GnWn+SOSEX)@PBE is excellent, in line with the findings reported in Ref.~\cite{Knight/etal:2016}. Remarkably, \GnWnG@PBE performs even better, with an MSE of only $-0.01$ eV and a MAE of 0.06 eV. For PBE0 and HF references, the performance of both methods noticeably deteriorates, but in each case, \GnWnG always performs slightly better than \GnWn+SOSEX. In contrast to the IPs, the EA deviations to the CCSD(T) reference data scatter very little.  In particular for the PBE starting point, the EA errors of both \GnWn+SOSEX and \GnWnG fall within the energy window of $[-0.2,0.2]$ eV almost perfectly for all of the 24 molecules. 

\begin{table*}[!ht]
\caption{\label{tab:Acceptor24_IPsEAs_MDs} MSDs (left side) and MADs (right side) of vertical IPs and EAs for the Acceptor24 test set 
as obtained by \GnWn, \GnWn+SOSEX and \GnWnG with respect to the CCSD(T) data~\cite{Richard2016/10.1021/acs.jctc.5b00875}. For each method, three
starting points (SP) -- PBE, PBE0, and HF are used.}
\begin{ruledtabular}
\begin{tabular}{l@{\hskip-10pt}    d@{\hskip-4pt}d  @{\hskip5pt}  d@{\hskip-4pt}d  @{\hskip10pt}  d@{\hskip-4pt}d    d    d@{\hskip-4pt}d  @{\hskip5pt}  d@{\hskip-4pt}d  @{\hskip10pt}  d@{\hskip-4pt}d}
&\multicolumn{6}{c}{IPs} & &\multicolumn{6}{c}{EAs} \\
\colrule
\multicolumn{1}{l}{\diagbox[width=1.9cm, height=0.78cm]{\raisebox{-0.3\height}{SP}}{Method}}
&\multicolumn{2}{c}{\textrm{$G_0W_0$}}     &\multicolumn{2}{c}{\textrm{$G_0W_0$+SOSEX}}     
&\multicolumn{2}{c}{\textrm{$G_0W_0\Gamma^{(1)}_0$}}
&
&\multicolumn{2}{c}{\textrm{$G_0W_0$}}     &\multicolumn{2}{c}{\textrm{$G_0W_0$+SOSEX}}     
&\multicolumn{2}{c}{\textrm{$G_0W_0\Gamma^{(1)}_0$}}   \\
\colrule
@PBE   &\quad-0.66&0.66   &\enspace\,-0.31&0.33   &\,-0.25&0.28   &   &\quad0.60&0.60   &\quad-0.05&0.08   &\enspace\,-0.01&0.06  \\
@PBE0  &\quad-0.21&0.22   &\enspace\,-0.06&0.12   &\, 0.16&0.16   &   &\quad0.39&0.39   &\quad-0.14&0.16   &\enspace\,-0.07&0.09  \\
@HF    &\quad 0.49&0.49   &\enspace\, 0.48&0.48   &\, 0.68&0.68   &   &\quad0.15&0.15   &\quad-0.20&0.22   &\enspace\,-0.13&0.17  \\
\end{tabular}
\end{ruledtabular}
\end{table*}

Finally, we briefly comment on the starting point dependence (SPD) of the \GnWn and beyond-\GnWn approaches. To quantitatively assess 
the SPD, we follow Ref.~\cite{Marom/etal:2012} and adopt the following parameter 
\begin{align}
\Delta_{\text{SPD}} &= \sum^N_{i=1} \frac{ | \epsilon^{\text{QP}}_{i,\text{HF}} - \epsilon^{\text{QP}}_{i,\text{PBE}}| }{N}
\label{eq:spd}
\end{align}
as an indicator to measure such a dependence. Here, $i$ runs over all the molecules contained in a given test set. 
Namely, for each method, the SPD is given by an average over the differences of the obtained QPEs
based on the HF and PBE starting points. In Table~\ref{tab:SPD}, we present the $\Delta_{\text{SPD}}$ values separately for
the IPs and EAs of the $GW100$ and Acceptor24 test sets. We observe that the SPD is reduced drastically for the $GW100$ IPs and the Acceptor24 IPs and EAs when going from \GnWn to \GnWn+SOSEX and \GnWnG.

\begin{table}[!ht]
\caption{\label{tab:SPD} The starting point dependence ($\Delta_{\text{SPD}}$ defined in Eq.~\ref{eq:spd}) for the IPs and EAs for the $GW100$ and Acceptor24 test sets}
\begin{ruledtabular}
\begin{tabular}{lcccc}
&\multicolumn{2}{c}{$GW100$}&\multicolumn{2}{c}{Acceptor24} \\
\cmidrule(lr){2-3}\cmidrule(lr){4-5}
\textrm{Method}&
\multicolumn{1}{c}{IP}&
\multicolumn{1}{c}{EA} &\multicolumn{1}{c}{IP}&\multicolumn{1}{c}{EA} \\
\colrule
$G_0W_0$                 &0.92  &0.22  &1.11  &0.45 \\
$G_0W_0$+SOSEX           &0.59  &0.23  &0.74  &0.15 \\
$G_0W_0\Gamma^{(1)}_0$   &0.68  &0.14  &0.89  &0.16 \\
\end{tabular}
\end{ruledtabular}
\end{table}

\subsection{Screening and the static approximation: The example of CO}

\begin{figure*}[!ht]
\includegraphics[scale=1]{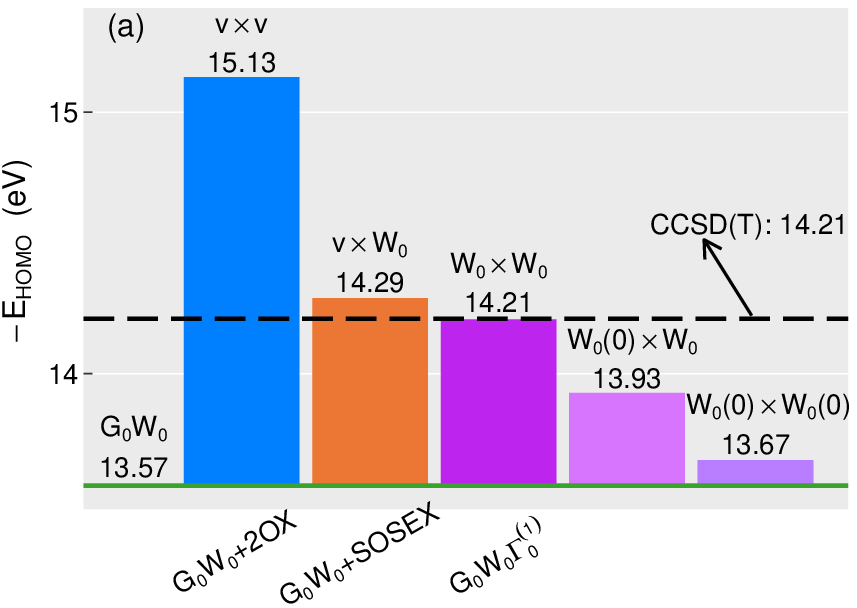}
\includegraphics[scale=1]{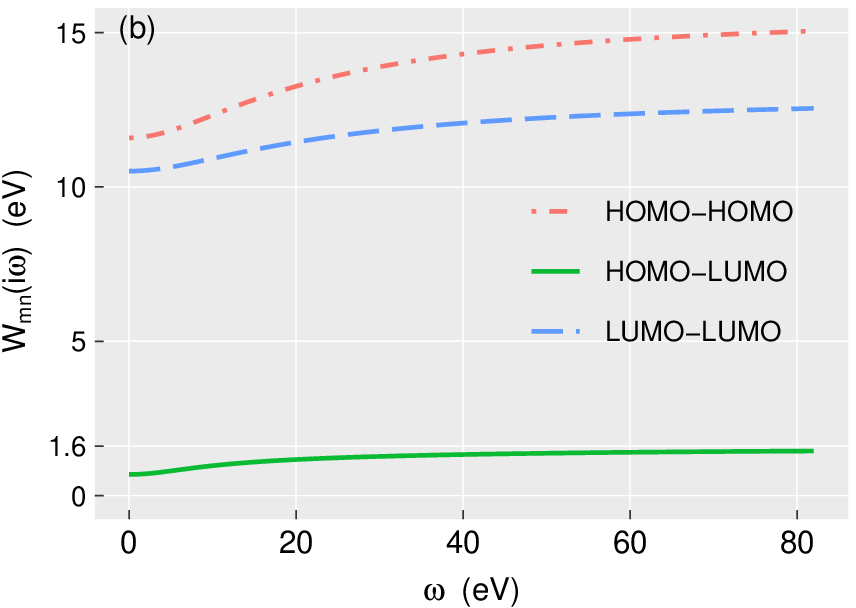}
\caption{
(a)  IP values of CO obtained with \GnWn and various beyond-\GnWn schemes based on PBE. The height of the colored bars indicates the difference to \GnWn. (b) The matrix elements of the screened Coulomb interaction $W_{0,mn}(i\omega)$ between the HOMO state (red, $m$ and $n$ being both HOMO),  the LUMO state (blue, $m$ and $n$ being both HOMO) are plotted as a function of imaginary frequency. The green line shows the off-diagonal matrix element between the HOMO and LUMO states ($m$ and $n$ being HOMO and LUMO, respectively). The calculations were performed for the CO bond length of 1.283~$\AA$ as used in the $GW$100 test set.  
}
\label{fig:CO_HOMO}
\end{figure*}

From the above benchmark tests for the $GW$100 and Acceptor24 molecules, it is evident that the
SOSEX and the FSOS-$W_0$  self-energies  have a significant affect on the QP energies. With rare exceptions (see the example of the benzene molecule below in Sec.~\ref{sec:benzene}), the magnitude of the two types of vertex corrections is similar in the sense that their differences are much smaller than the corrections themselves. This observation is surprising since FSOS-$W_0$ involves two screened Coulomb lines in contrast to SOSEX whose second interaction is bare. One would thus expect certain quantitative differences between $GW$+SOSEX and \GnWnG.

To have a better idea of what is happening, we present in Fig.~\ref{fig:CO_HOMO} the IP values of CO for \GnWn, \GnWn+2OX, \GnWn+SOSEX, and \GnWnG based on the PBE starting point. The height of the colored bars in Fig.~\ref{fig:CO_HOMO} is referenced to the \GnWn@PBE value of 13.57~eV. We observe that the 2OX term gives rise to the largest change, pushing the CO IP from 13.57~eV to 15.13~eV. When one Coulomb line gets screened in the SOSEX self-energy the IP drops to 14.29~eV. This screening thus reduces the self-energy correction by a factor of 2 from 1.56~eV for 2OX to 0.72~eV for SOSEX. Further screening by the 2nd $W$ in FSOS-$W_0$ has only a comparatively small contribution reducing the self-energy correction to 0.64~eV. The resulting \GnWnG@PBE IP  of 14.21 eV  is in excellent agreement with the CCSD(T) reference result. 

Our analysis suggests that the effect of screening the second Coulomb line (i.e., from SOSEX to FSOS-$W_0$) is one order of magnitude smaller than screening the first Coulomb line (i.e., from 2OX to SOSEX). The overall difference between \GnWn+SOSEX and \GnWnG is of the order of 0.1~eV,  which is much smaller than the SOSEX or FSOS-$W_0$ corrections themselves. This observation is consistent with the overall behavior of \GnWn+SOSEX and \GnWnG for the $GW$100 and Acceptor24 test sets.

Also presented in Fig.~\ref{fig:CO_HOMO} are the results obtained by two variants of the \GnWnG scheme, in which either one or both of the frequency-dependent screened Coulomb interactions are replaced by their value at zero frequency, i.e., $W_0(i\omega) \approx W(0)$. The fully static approximation is common in the literature \cite{Bobbert/Haeringen:1994,Grueneis/etal:2014}, because it computationally significantly more efficient. However, the static approximation has a drastic effect on the \GnWnG self-energy correction as can be seen clearly in Fig.~\ref{fig:CO_HOMO}. Making one screened Coulomb line static
[denoted $W_0(0)\times W_0$ in Fig.~\ref{fig:CO_HOMO}(a)] reduces the magnitude of the FSOS-$W_0$ correction by a factor of 2. Making also the 2nd $W_0$ static [denoted $W_0(0)\times W_0(0)$ in Fig.~\ref{fig:CO_HOMO}(a)] reduces the self-energy correction even further. The CO IP now amounts to only 13.67~eV and is just 0.1~eV larger than that in $G_0W_0$. 

The behaviour of the static approximation can be understood from the frequency dependency of the screened Coulomb interaction. Figure~\ref{fig:CO_HOMO}(b) presents the matrix elements of the screened Coulomb interaction $W_{0,mn}(i\omega)=\langle mn|W_0(i\omega)|mn \rangle$ as a function of imaginary frequency. The three lines plotted in Fig.~\ref{fig:CO_HOMO}(b) correspond to the matrix elements associated purely with the
HOMO state ($m=n=n_\text{HOMO}$), the LUMO state ($m=n=n_\text{LUMO}$), and with both the HOMO and
LUMO states ($m=n_\text{HOMO},n=n_\text{LUMO}$). 
The plots show that the screened Coulomb interaction at the
zero frequency is the minimum along the imaginary frequency axis. Thus the screened Coulomb interaction at the zero
frequency is a rather poor representation of its average strength over the entire frequency range, and will necessarily 
lead to a substantial underestimation of the vertex correction.

\begin{figure}[!ht]
\includegraphics[scale=1]{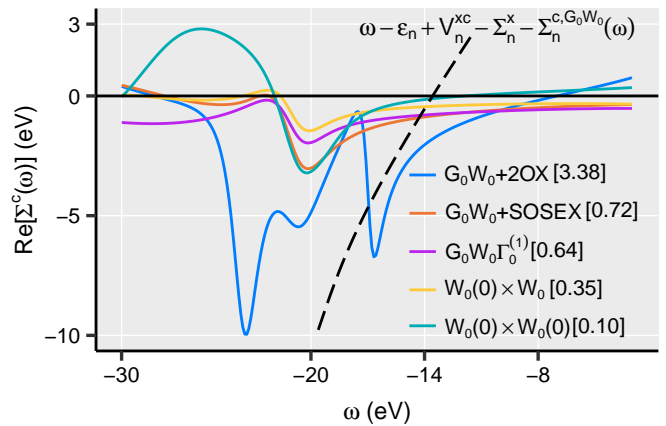}
\caption{ The beyond-\GnWn parts of the correlation self-energies on real-frequency axis obtained with five beyond-\GnWn schemes.
The dashed line represents $f(\omega)=\omega-\epsilon_n+V_n^\text{xc}-\Sigma^\text{x}_n - \Sigma^\text{c,\GnWn}_n(\omega)$, with 
$\epsilon_n$, $V_n^\text{xc}$, $\Sigma^\text{x}_n$ and $\Sigma^\text{c,\GnWn}_n(\omega)$  being the KS eigenvalues, the diagonal matrix elements of the PBE-XC potential, and the exact-exchange and \GnWn correlation self-energy, respectively. Its intersections with the self-energy lines yield the
contributions of the vertex correction to the QPE. The values [in eV] of the beyond-\GnWn corrections are given in the bracket 
behind the legends of individual computational schemes. 
}
\label{fig:CO_HOMO_selfenergy}
\end{figure}

To gain further insights into the effect brought by the static approximation of the screened Coulomb interaction, in Fig.~\ref{fig:CO_HOMO_selfenergy} 
we present the beyond-\GnWn self-energy corrections of the CO HOMO level on real frequency axes, 
from which the influences of different treatments of the screened Coulomb interaction on the self-energies are clearly visualized. 
It should be noted that even the (bare or screened) interaction is static, the resultant self-energy at second or higher order is still
frequency dependent, due to the presence of multiple Green functions.
From Fig.~\ref{fig:CO_HOMO_selfenergy}, one may fist notice that the bare 2OX part of the self-energy is drastically
different from other schemes, in which one or both of the Coulomb lines are screened. In fact, the 2OX correction is so different from the \GnWn self-energy
itself such that a self-consistent solution of Eq.~\ref{eq:qp_eqn} cannot be found iteratively for the \GnWn+2OX self-energy. The \GnWn+2OX QP energy
for CO HOMO reported in Fig.~\ref{fig:CO_HOMO}(a) was found by linearizing the \GnWn+2OX self-energy around the KS eigenvalue, which differs significantly
from the graphical solution indicated in Fig.~\ref{fig:CO_HOMO}(c). The self-energy corrections provided by other beyond-\GnWn schemes are much milder and
their magnitudes around the QPE of CO HOMO follows the sequence of $W_0(0)\times W_0(0)$, $W_0(0)\times W_0$, FSOS-W$_0$, and SOSEX. The corrections
to the \GnWn QPE of CO HOMO level are given by the intersections in Fig.~\ref{fig:CO_HOMO}(c), which are consistent with IP values obtained by these
beyond-\GnWn schemes reported in Fig.~\ref{fig:CO_HOMO}(a).

\subsection{\label{sec:benzene}Photoemission spectra of the benzene molecule}

Next we revisit a problem in the photoemission spectrum  (PES) of benzene (C$_6$H$_6$) we observed in Ref.~\cite{Ren/etal:2015}. Benzene is a relatively
simple molecule, but the relative energy between the HOMO-1 (with $e_{2g}(\sigma)$ character) and the HOMO-2 
(with $a_{2u}(\pi)$ character) states is poorly  described by \GnWn irrespective of the starting point and also by self-consistent $GW$ methods. We observed in our previous work \cite{Ren/etal:2015} that the energy spacing between these two states is controlled by higher-order exchange effects absent from $GW$. The bare or screened second-order exchange terms restore correct the energy ordering of the HOMO-1 and HOMO-2 states, but the energy splitting between the two states is still overestimated \cite{Ren/etal:2015}.

\begin{figure}[!ht]
\includegraphics[scale=1]{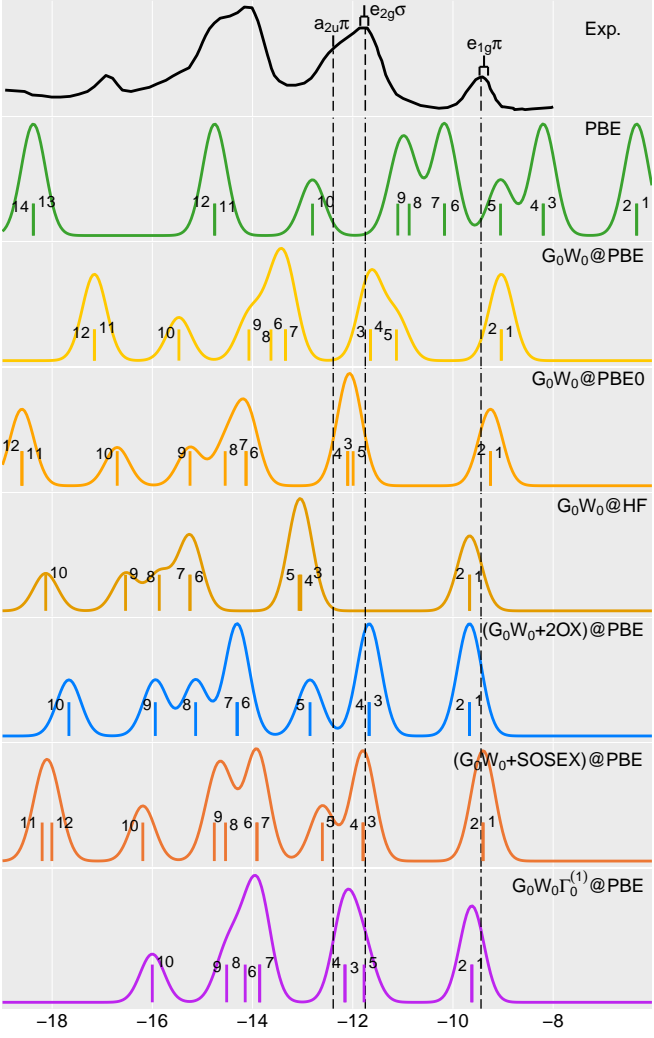}
\caption{The valence QP density of states of the benzene molecule for different \GnWn and beyond-\GnWn approximations. The numbers label the original order of the peaks that appears in the PBE calculations, starting with the (two-fold degenerate) HOMO level marked as ``1 2". The experimental photoemission spectrum of Liu \textit{et al.} \cite{Liu/etal:2011} is included for comparison.}
\label{fig:benzene_spectrum}
\end{figure}

In Fig.~\ref{fig:benzene_spectrum}, we present the QP density of states (DOS) for different approaches, in comparison to the experimental PES spectrum \cite{Liu/etal:2011}. The theoretical spectra are artificially broadened by 0.24~eV. In both PBE and experiment, the non-degenerate $a_{2u}(\pi)$ state (labelled ``5" in Fig.~\ref{fig:benzene_spectrum}) falls well below the  2-fold degenerate $e_{2g}(\sigma)$ state (labelled ``4 3" in Fig.~\ref{fig:benzene_spectrum}). However, for \GnWn case, regardless of the starting point, the $a_{2u}(\pi)$ state is either above or energetically very close to the $e_{2g}(\sigma)$ state. Note that the \GnWn results presented in Fig.~\ref{fig:benzene_spectrum} 
is appreciably different from those in Fig.~3 of Ref.~\cite{Ren/etal:2015}. For example, for \GnWn@PBE, as shown in
Fig.~\ref{fig:benzene_spectrum} of the present work, the computed
$e_{2g}(\sigma)$ is clearly below the $a_{2u}(\pi)$ state, whereas it
is still slightly above $a_{2u}(\pi)$ in Ref.~\cite{Ren/etal:2015}. This discrepancy is primarily due to the different
analytical continuation methods used in the two works: In Ref.~\cite{Ren/etal:2015} the simple two-pole fitting procedure
was used and not the the Pad\'e approximation as  adopted in the present work. It is striking that, compared to the
Pad\'e result, the $a_{2u}(\pi)$ level is shifted downwards by as much as 0.5 - 1 eV by the two-pole fitting scheme.
Benchmark calculations at the \GnWn level against the reference results obtained using the contour deformation scheme
\cite{Golze/etal:2018} indicate that the Pad\'e approximation is very accurate whereas the two-pole fitting procedure is not
reliable in the present case. Hence all the \GnWn and beyond-\GnWn results presented in Fig.~\ref{fig:benzene_spectrum} are obtained
using the Pad\'e approximation.

Consistent with Ref.~\cite{Ren/etal:2015}, (\GnWn+2OX)@PBE and (\GnWn+SOSEX)@PBE restore the correct energy ordering between $a_{2u}(\pi)$ and $e_{2g}(\sigma)$. However, it is somewhat disappointing that \GnWnG does not produce the correct energy ordering. Compared to \GnWn@PBE, both the $e_{2g}(\sigma)$ and the $a_{2u}(\pi)$ states are correctly shifted to higher binding energies in \GnWnG@PBE, but the relative energy ordering of the two states is not reversed. So far, this is the only case that we found, in which \GnWnG yields qualitatively different (and unfortunately incorrect) results compared to \GnWn+SOSEX. At present, we have not yet fully understood this behavior. As a summary, in Table~\ref{tab:benzene_split}, we present the energy separations between the HOMO-1 and HOMO-2 states of benzene, as obtained by various approaches. The results reported in the literature,
obtained by dynamical configuration interaction (DCI)~\cite{Dvorak/Golze/Rinke:2019}, renormalized single $GW$ \cite{Jin/Su/Yang:2019}, and IP-EOM-CCSD \cite{Ranasinghe/etal:2019} are also included here for comparison. It appears
that ($G_0W_0$+SOSEX)@PBE and IP-EOM-CCSD are the only two approaches examined here which can produce satisfactory energy
separations of the HOMO-1 and HOMO-2 states of benzene.
\begin{table}[!ht]
\caption{\label{tab:benzene_split}QPE levels of the HOMO-1 and HOMO-2 states, as well as their separations, as calculated 
by various theoretical approaches, compared to the experimental value. A negative ``splitting" value indicates that
the HOMO-1 and HOMO-2 levels are erroneously reversed. The DCI, RSGW, and IP-EOM-CCSD results are taken
from the literature.}
\begin{ruledtabular}
\begin{tabular}{lddd}
\textrm{Method}&
\multicolumn{1}{c}{$-\text{E}_{\text{HOMO}-1}$}&
\multicolumn{1}{c}{$-\text{E}_{\text{HOMO}-2}$}&
\multicolumn{1}{c}{Splitting}\\
\colrule
Exp.                &11.75   &12.39   & 0.64 \\
PBE                 & 8.20   & 9.05   & 0.85 \\
$G_0W_0$@PBE        &11.65   &11.13   &-0.52 \\
$G_0W_0$@PBE0       &12.10   &11.99   &-0.11 \\
$G_0W_0$@HF         &13.04   &13.07   & 0.03 \\
($G_0W_0$+2OX)@PBE    &11.67   &12.85   & 1.18 \\
($G_0W_0$+SOSEX)@PBE  &11.79   &12.61   & 0.81 \\
$G_0W_0\Gamma^{(1)}_0$@PBE  &12.16   &11.77   &-0.38 \\
DCI$^a$                     & 12.28 & 12.45      & 0.17 \\
RSGW@PBE$^b$                 &        &      & 0.1 \\
IP-EOM-CCSD$^c$                 &   12.18  & 12.65  & 0.47
\end{tabular}
\end{ruledtabular}
\begin{tabular}{lll}
$^a$Ref~\cite{Dvorak/Golze/Rinke:2019}~~~~~~ & ~~~~~~$^b$Ref~\cite{Jin/Su/Yang:2019}~~~~~~ & ~~~~~~$^c$Ref~\cite{Ranasinghe/etal:2019}
\end{tabular}
\end{table}

\section{\label{sec:conclu-outlook} Conclusion and Outlook}

In this work, we implemented the one-shot version of the \GWG scheme (\GnWnG) in an atomic-orbital basis set framework. \GWG  goes beyond $GW$ by including the full second-order self-energy within Hedin's formalism. The formal scaling of the \GnWnG scheme is $O(N^5)$, but we achieved an improved scaling of $O(N^{4.2})$ for medium-sized molecules.

Benchmarks for the $GW$100 and Acceptor24 test sets reveal that \GnWnG significantly outperforms \GnWn for PBE and PBE0 starting points, whereas HF is not a good starting point for \GnWnG. We further observe that the accuracy of \GnWnG is similar to the previously developed \GnWn+SOSEX approach despite their different diagrammatic structure. We also checked the influence of static screened Coulomb approximations, and found that they significantly underestimate the magnitude of the vertex corrections.

Compared to $GW$+SOSEX, \GWG is a conserving approximation in the sense that its self-energy diagrams can be derived from an associated $\Psi$-diagram \cite{Baym/Kadanoff:1961,Almbladh/etal:1999,Dahlen/Leeuwen/Barth:2006}. Thus a meaningful self-consistent \GWG scheme can be defined, similar to the $GW$ case. However, the behavior of such a self-consistent \GWG scheme, in particular, how the non-positiveness of the spectral function observed in jellium \cite{Unimonen/vanLeeuwen:2014,Stefanucci/vanLeeuwen:2015} is manifested in molecular systems, remains to be studied.

\begin{acknowledgments}
We thank Marc Dvorak for sharing his DCI data for the benzene molecule. 
This work was funded by the National Key Research and Development Program of China (Grants No. 2016YFB0201202), 
the Chinese National Science Foundation Grant number 11874335, and the Max Planck Partner Group project 
on \textit{Advanced Electronic Structure Methods}.
\end{acknowledgments}

\begin{appendix}

\section{\label{appendix:diagram} Diagrammatic representation of the self-energy expansion up to the 3rd order in $W$}

\begin{figure}
    \centering
    \includegraphics[scale=0.5]{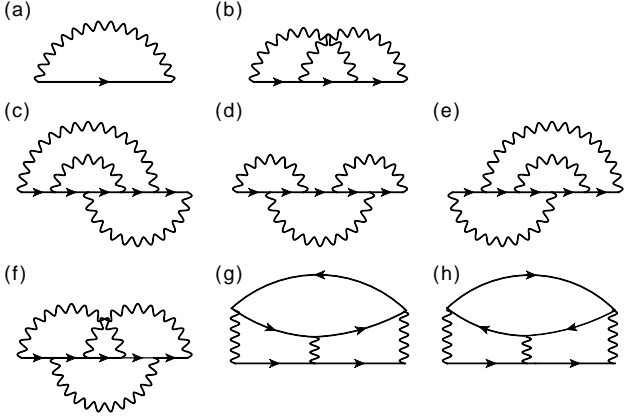}
    \caption{Self-energy diagrams up to third order in $W$: (a) the first-order $GW$ diagram,
     (b) the second-order FSOS-$W$ diagram, (c)-(h)  third-order diagrams.}
    \label{fig:app_diagrams}
\end{figure}

The diagrams of the  self-energy expansion up the third order in the screened Coulomb interaction $W$ are
presented in Fig.~\ref{fig:app_diagrams}. In the present work, only contributions from the first two orders are included in $\GnWnG$ calculations.

\section{\label{appendix:implemn} Implementation algorithm for the FSOS-$W_0$ self-energy}

\begin{figure*}
\centering
\includegraphics{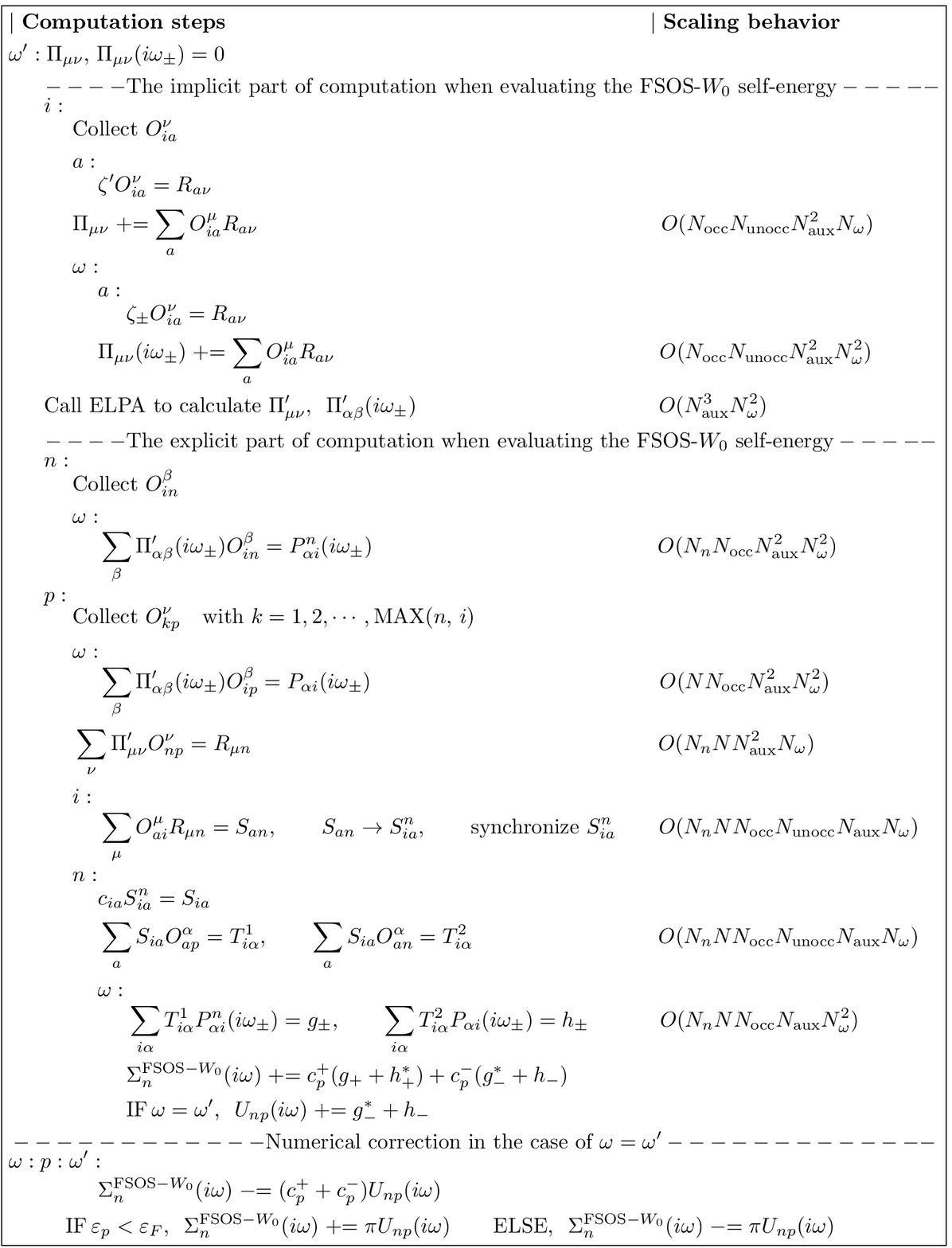}
\caption{\label{fig:fsos_algom} The implemented algorithm to compute the FSOS-$W_0$ self-energy following Eq.(\ref{eq:FSOS_implemn}).
Note that, ($\nu,\,\beta$) and ($\alpha,\,\mu$) are global and local indices respectively, regarding the parallelization scheme. The additional $U$-matrix correction at the bottom is intended to remove numerical errors by analytic solution when the two frequency values are equal. Scaling properties of essential operations with respect to the number of states and frequency points are 
indicated at the end of each line. For simplicity, the following abbreviations have been used: 
$\Pi'=(1- \Pi)^{-1}, \, \zeta'=2(\varepsilon_i - \varepsilon_a)/[ (\varepsilon_i - \varepsilon_a)^2 +\omega'^2],\, \zeta_{\pm}=2(\varepsilon_i - \varepsilon_a)/[ (\varepsilon_i - \varepsilon_a)^2 
+ (\omega\pm\omega')^2],\, c_{ia} = 1/\left( i\omega' + \varepsilon_i - \varepsilon_a \right)$ 
and $c^{\pm}_p=1/\left( i\omega\pm i\omega'+\varepsilon_{\text{F}}-\varepsilon_p \right)$.}
\end{figure*}

Here we present the loop structure and the parallelization scheme for the FSOS-$W_0$ self-energy as implemented in FHI-aims. To start with, we rewrite Eq.~\ref{eq:FSOS_implemn} as,
\begin{widetext}
\begin{eqnarray}
&&\Sigma^{ \text{\scriptsize FSOS} }_n = \int^{\infty}_0 \frac{\text{d}\omega'}{2\pi}\, 
\sum_p \sum^{\text{occ}}_i \sum^{\text{unocc}}_a \sum_{\mu\nu} 
O^{\mu}_{np}\, [\,1- \Pi(i\omega') \,]^{-1}_{\mu\nu} \, O^{\nu}_{ia} \times \nonumber \\
&&\qquad\quad \sum_{\alpha\beta} \Bigg\lbrace \enspace O^{\alpha}_{ap} 
\left[ \frac{ [\, 1- \Pi(i\omega + i\omega') \,]^{-1}_{\alpha\beta} \,O^{\beta}_{in} }
{ (i\omega + i\omega' + \varepsilon_{\text{F}} - \varepsilon_p)(i\omega' + \varepsilon_i 
- \varepsilon_a) } - \frac{ [\, 1- \Pi(i\omega - i\omega') \,]^{-1}_{\alpha\beta} \, O^{\beta}_{in} }
{ (i\omega - i\omega' + \varepsilon_{\text{F}} -\varepsilon_p)(i\omega' + \varepsilon_a 
- \varepsilon_i) } \right] \nonumber \\
&&\qquad\qquad\qquad 
-O^{\alpha}_{an} \left[ \frac{ [\, 1- \Pi(i\omega + i\omega') \,]^{-1}_{\alpha\beta}\, O^{\beta}_{ip} }
{ (i\omega + i\omega' + \varepsilon_{\text{F}} - \varepsilon_p)(i\omega' + \varepsilon_a 
- \varepsilon_i) } - \frac{ [\, 1- \Pi(i\omega - i\omega') \,]^{-1}_{\alpha\beta} \, O^{\beta}_{ip} }
{ (i\omega - i\omega' +\varepsilon_{\text{F}} - \varepsilon_p)(i\omega' + \varepsilon_i 
- \varepsilon_a) } \right] \, \Bigg\rbrace. \label{FSOS-implement}
\label{eq:app:fsos-w0}
\end{eqnarray}
\end{widetext}
whereby integer occupancy for the MOs is assumed.
Our actual implementation, which is directly based on Eq.~\ref{eq:app:fsos-w0}, is outlined in the algorithm presented in Fig.~\ref{fig:fsos_algom}.

We will not go into the details of the algorithm, but only briefly explain the main considerations behind this particular loop structure, which we found by extensive testing of different choices. The key point is to make the frequency $\omega'$  the outermost variable of the loop structure, because $\omega'$ is present in all matrix multiplications. We therefore gain computational and memory efficiency. Within the $\omega'$ loop, we first prepare intermediate quantities like the $\Pi$ matrices at two different frequencies before looping over the orbital index $p$. The orbital $p$ is special compared to $i$ and $a$, as it runs over both occupied and unoccupied states and couples to frequency $\omega$. A further reduction of the prefactor can be gained from a careful arrangement of the orbital indices $i$ and $a$ since the number of unoccupied states is typically much larger than that of the occupied ones. 

Despite the algorithmic optimization, the evaluation of the FSOS-W$_0$ self-energy is still computationally expensive. To reduce computational load and memory storage, we need an efficient parallelization. The key part of our paralleization is the treatment of the 3-dimensional $O$ array, which is distributed with Message Passing  Interface (MPI) processes across the auxiliary basis index $\nu$, and needs to be collected when computing the $\Pi$ matrices. The essential point is to ensure that the computationally heavy parts of the calculation, especially those inside deeply-nested loops, are parallelized. As can be seen in in Fig.~\ref{fig:fsos_algom}, all the heavy computation steps are parallelized except for several unavoidable light  product operations that will not affect the overall performance.

\end{appendix}
\bibliography{CommonBib}

\begin{thebibliography}{93}%
\makeatletter
\providecommand \@ifxundefined [1]{%
 \@ifx{#1\undefined}
}%
\providecommand \@ifnum [1]{%
 \ifnum #1\expandafter \@firstoftwo
 \else \expandafter \@secondoftwo
 \fi
}%
\providecommand \@ifx [1]{%
 \ifx #1\expandafter \@firstoftwo
 \else \expandafter \@secondoftwo
 \fi
}%
\providecommand \natexlab [1]{#1}%
\providecommand \enquote  [1]{``#1''}%
\providecommand \bibnamefont  [1]{#1}%
\providecommand \bibfnamefont [1]{#1}%
\providecommand \citenamefont [1]{#1}%
\providecommand \href@noop [0]{\@secondoftwo}%
\providecommand \href [0]{\begingroup \@sanitize@url \@href}%
\providecommand \@href[1]{\@@startlink{#1}\@@href}%
\providecommand \@@href[1]{\endgroup#1\@@endlink}%
\providecommand \@sanitize@url [0]{\catcode `\\12\catcode `\$12\catcode
  `\&12\catcode `\#12\catcode `\^12\catcode `\_12\catcode `\%12\relax}%
\providecommand \@@startlink[1]{}%
\providecommand \@@endlink[0]{}%
\providecommand \url  [0]{\begingroup\@sanitize@url \@url }%
\providecommand \@url [1]{\endgroup\@href {#1}{\urlprefix }}%
\providecommand \urlprefix  [0]{URL }%
\providecommand \Eprint [0]{\href }%
\providecommand \doibase [0]{https://doi.org/}%
\providecommand \selectlanguage [0]{\@gobble}%
\providecommand \bibinfo  [0]{\@secondoftwo}%
\providecommand \bibfield  [0]{\@secondoftwo}%
\providecommand \translation [1]{[#1]}%
\providecommand \BibitemOpen [0]{}%
\providecommand \bibitemStop [0]{}%
\providecommand \bibitemNoStop [0]{.\EOS\space}%
\providecommand \EOS [0]{\spacefactor3000\relax}%
\providecommand \BibitemShut  [1]{\csname bibitem#1\endcsname}%
\let\auto@bib@innerbib\@empty
\bibitem [{\citenamefont {Fetter}\ and\ \citenamefont
  {Walecka}(1971)}]{Fetter/Walecka:1971}%
  \BibitemOpen
  \bibfield  {author} {\bibinfo {author} {\bibfnamefont {A.~L.}\ \bibnamefont
  {Fetter}}\ and\ \bibinfo {author} {\bibfnamefont {J.~D.}\ \bibnamefont
  {Walecka}},\ }\href@noop {} {\emph {\bibinfo {title} {Quantum Theory of
  Many-Particle Systems}}}\ (\bibinfo  {publisher} {McGraw-Hill},\ \bibinfo
  {address} {New York},\ \bibinfo {year} {1971})\BibitemShut {NoStop}%
\bibitem [{\citenamefont {Martin}\ \emph {et~al.}(2016)\citenamefont {Martin},
  \citenamefont {Reining},\ and\ \citenamefont
  {Ceperley}}]{martin_reining_ceperley_2016}%
  \BibitemOpen
  \bibfield  {author} {\bibinfo {author} {\bibfnamefont {R.~M.}\ \bibnamefont
  {Martin}}, \bibinfo {author} {\bibfnamefont {L.}~\bibnamefont {Reining}},\
  and\ \bibinfo {author} {\bibfnamefont {D.~M.}\ \bibnamefont {Ceperley}},\
  }\href {https://doi.org/10.1017/CBO9781139050807} {\emph {\bibinfo {title}
  {Interacting Electrons: Theory and Computational Approaches}}}\ (\bibinfo
  {publisher} {Cambridge University Press},\ \bibinfo {year}
  {2016})\BibitemShut {NoStop}%
\bibitem [{\citenamefont {Hedin}(1965)}]{Hedin:1965}%
  \BibitemOpen
  \bibfield  {author} {\bibinfo {author} {\bibfnamefont {L.}~\bibnamefont
  {Hedin}},\ }\href@noop {} {\bibfield  {journal} {\bibinfo  {journal} {Phys.\
  Rev.}\ }\textbf {\bibinfo {volume} {139}},\ \bibinfo {pages} {A796} (\bibinfo
  {year} {1965})}\BibitemShut {NoStop}%
\bibitem [{\citenamefont {Hybertsen}\ and\ \citenamefont
  {Louie}(1986)}]{Hybertsen/Louie:1986}%
  \BibitemOpen
  \bibfield  {author} {\bibinfo {author} {\bibfnamefont {M.~S.}\ \bibnamefont
  {Hybertsen}}\ and\ \bibinfo {author} {\bibfnamefont {S.~G.}\ \bibnamefont
  {Louie}},\ }\href@noop {} {\bibfield  {journal} {\bibinfo  {journal} {Phys.
  Rev. B}\ }\textbf {\bibinfo {volume} {34}},\ \bibinfo {pages} {5390}
  (\bibinfo {year} {1986})}\BibitemShut {NoStop}%
\bibitem [{\citenamefont {Godby}\ \emph {et~al.}(1986)\citenamefont {Godby},
  \citenamefont {Schl\"uter},\ and\ \citenamefont
  {Sham}}]{Godby/Schlueter/Sham:1986}%
  \BibitemOpen
  \bibfield  {author} {\bibinfo {author} {\bibfnamefont {R.~W.}\ \bibnamefont
  {Godby}}, \bibinfo {author} {\bibfnamefont {M.}~\bibnamefont {Schl\"uter}},\
  and\ \bibinfo {author} {\bibfnamefont {L.~J.}\ \bibnamefont {Sham}},\
  }\href@noop {} {\bibfield  {journal} {\bibinfo  {journal} {Phys. Rev. Lett.}\
  }\textbf {\bibinfo {volume} {56}},\ \bibinfo {pages} {2415} (\bibinfo {year}
  {1986})}\BibitemShut {NoStop}%
\bibitem [{\citenamefont {Aryasetiawan}\ and\ \citenamefont
  {Gunnarsson}(1998)}]{Aryasetiawan/Gunnarsson:1998}%
  \BibitemOpen
  \bibfield  {author} {\bibinfo {author} {\bibfnamefont {F.}~\bibnamefont
  {Aryasetiawan}}\ and\ \bibinfo {author} {\bibfnamefont {O.}~\bibnamefont
  {Gunnarsson}},\ }\href {https://doi.org/10.1088/0034-4885/61/3/002}
  {\bibfield  {journal} {\bibinfo  {journal} {Rep. Prog. Phys.}\ }\textbf
  {\bibinfo {volume} {61}},\ \bibinfo {pages} {237} (\bibinfo {year}
  {1998})}\BibitemShut {NoStop}%
\bibitem [{\citenamefont {Rinke}\ \emph {et~al.}(2005)\citenamefont {Rinke},
  \citenamefont {Qteish}, \citenamefont {Neugebauer}, \citenamefont
  {Freysoldt},\ and\ \citenamefont {Scheffler}}]{Rinke/etal:2005}%
  \BibitemOpen
  \bibfield  {author} {\bibinfo {author} {\bibfnamefont {P.}~\bibnamefont
  {Rinke}}, \bibinfo {author} {\bibfnamefont {A.}~\bibnamefont {Qteish}},
  \bibinfo {author} {\bibfnamefont {J.}~\bibnamefont {Neugebauer}}, \bibinfo
  {author} {\bibfnamefont {C.}~\bibnamefont {Freysoldt}},\ and\ \bibinfo
  {author} {\bibfnamefont {M.}~\bibnamefont {Scheffler}},\ }\href@noop {}
  {\bibfield  {journal} {\bibinfo  {journal} {New J.\ Phys.}\ }\textbf
  {\bibinfo {volume} {7}},\ \bibinfo {pages} {126} (\bibinfo {year}
  {2005})}\BibitemShut {NoStop}%
\bibitem [{\citenamefont {\mbox{van Schilfgaarde}}\ \emph
  {et~al.}(2006)\citenamefont {\mbox{van Schilfgaarde}}, \citenamefont
  {Kotani},\ and\ \citenamefont {Faleev}}]{Schilfgaarde/Kotani/Faleev:2006}%
  \BibitemOpen
  \bibfield  {author} {\bibinfo {author} {\bibfnamefont {M.}~\bibnamefont
  {\mbox{van Schilfgaarde}}}, \bibinfo {author} {\bibfnamefont
  {T.}~\bibnamefont {Kotani}},\ and\ \bibinfo {author} {\bibfnamefont
  {S.}~\bibnamefont {Faleev}},\ }\href@noop {} {\bibfield  {journal} {\bibinfo
  {journal} {Phys. Rev. Lett.}\ }\textbf {\bibinfo {volume} {96}},\ \bibinfo
  {pages} {226402} (\bibinfo {year} {2006})}\BibitemShut {NoStop}%
\bibitem [{\citenamefont {Shishkin}\ \emph {et~al.}(2007)\citenamefont
  {Shishkin}, \citenamefont {Marsman},\ and\ \citenamefont
  {Kresse}}]{Shishkin/Marsman/Kresse:2007}%
  \BibitemOpen
  \bibfield  {author} {\bibinfo {author} {\bibfnamefont {M.}~\bibnamefont
  {Shishkin}}, \bibinfo {author} {\bibfnamefont {M.}~\bibnamefont {Marsman}},\
  and\ \bibinfo {author} {\bibfnamefont {G.}~\bibnamefont {Kresse}},\ }\href
  {https://doi.org/10.1103/PhysRevLett.99.246403} {\bibfield  {journal}
  {\bibinfo  {journal} {Phys. Rev. Lett.}\ }\textbf {\bibinfo {volume} {99}},\
  \bibinfo {pages} {246403} (\bibinfo {year} {2007})}\BibitemShut {NoStop}%
\bibitem [{\citenamefont {Gatti}\ \emph {et~al.}(2007)\citenamefont {Gatti},
  \citenamefont {Bruneval}, \citenamefont {Olevano},\ and\ \citenamefont
  {Reining}}]{Gatti/Bruneval/Olevano/Reining:2007}%
  \BibitemOpen
  \bibfield  {author} {\bibinfo {author} {\bibfnamefont {M.}~\bibnamefont
  {Gatti}}, \bibinfo {author} {\bibfnamefont {F.}~\bibnamefont {Bruneval}},
  \bibinfo {author} {\bibfnamefont {V.}~\bibnamefont {Olevano}},\ and\ \bibinfo
  {author} {\bibfnamefont {L.}~\bibnamefont {Reining}},\ }\href
  {https://doi.org/10.1103/PhysRevLett.99.266402} {\bibfield  {journal}
  {\bibinfo  {journal} {Phys. Rev. Lett.}\ }\textbf {\bibinfo {volume} {99}},\
  \bibinfo {pages} {266402} (\bibinfo {year} {2007})}\BibitemShut {NoStop}%
\bibitem [{\citenamefont {Friedrich}\ \emph {et~al.}(2010)\citenamefont
  {Friedrich}, \citenamefont {Bl\"{u}gel},\ and\ \citenamefont
  {Schindlmayr}}]{Friedrich/etal:2010}%
  \BibitemOpen
  \bibfield  {author} {\bibinfo {author} {\bibfnamefont {C.}~\bibnamefont
  {Friedrich}}, \bibinfo {author} {\bibfnamefont {S.}~\bibnamefont
  {Bl\"{u}gel}},\ and\ \bibinfo {author} {\bibfnamefont {A.}~\bibnamefont
  {Schindlmayr}},\ }\href@noop {} {\bibfield  {journal} {\bibinfo  {journal}
  {Phys. Rev. B}\ }\textbf {\bibinfo {volume} {81}},\ \bibinfo {pages} {125102}
  (\bibinfo {year} {2010})}\BibitemShut {NoStop}%
\bibitem [{\citenamefont {Jiang}\ and\ \citenamefont
  {Blaha}(2016)}]{Jiang/Blaha:2016}%
  \BibitemOpen
  \bibfield  {author} {\bibinfo {author} {\bibfnamefont {H.}~\bibnamefont
  {Jiang}}\ and\ \bibinfo {author} {\bibfnamefont {P.}~\bibnamefont {Blaha}},\
  }\href@noop {} {\bibfield  {journal} {\bibinfo  {journal} {Phys. Rev. B}\
  }\textbf {\bibinfo {volume} {93}},\ \bibinfo {pages} {115203} (\bibinfo
  {year} {2016})}\BibitemShut {NoStop}%
\bibitem [{\citenamefont {Golze}\ \emph {et~al.}(2019)\citenamefont {Golze},
  \citenamefont {Dvorak},\ and\ \citenamefont
  {Rinke}}]{Golze/Dvorak/Rinke:2019}%
  \BibitemOpen
  \bibfield  {author} {\bibinfo {author} {\bibfnamefont {D.}~\bibnamefont
  {Golze}}, \bibinfo {author} {\bibfnamefont {M.}~\bibnamefont {Dvorak}},\ and\
  \bibinfo {author} {\bibfnamefont {P.}~\bibnamefont {Rinke}},\ }\href@noop {}
  {\bibfield  {journal} {\bibinfo  {journal} {Frontiers in Chemistry}\ }\textbf
  {\bibinfo {volume} {7}},\ \bibinfo {pages} {377} (\bibinfo {year}
  {2019})}\BibitemShut {NoStop}%
\bibitem [{\citenamefont {Rangel}\ \emph {et~al.}(2020)\citenamefont {Rangel},
  \citenamefont {{Del Ben}}, \citenamefont {Varsano}, \citenamefont {Antonius},
  \citenamefont {Bruneval}, \citenamefont {{da Jornada}}, \citenamefont {J.{van
  Setten}}, \citenamefont {Orhan}, \citenamefont {{O'Regan}}, \citenamefont
  {Canning}, \citenamefont {Ferretti}, \citenamefont {Marini}, \citenamefont
  {Rignanese}, \citenamefont {Deslippe}, \citenamefont {Louie},\ and\
  \citenamefont {Neaton}}]{Rangel/etal:2020}%
  \BibitemOpen
  \bibfield  {author} {\bibinfo {author} {\bibfnamefont {T.}~\bibnamefont
  {Rangel}}, \bibinfo {author} {\bibfnamefont {M.}~\bibnamefont {{Del Ben}}},
  \bibinfo {author} {\bibfnamefont {D.}~\bibnamefont {Varsano}}, \bibinfo
  {author} {\bibfnamefont {G.}~\bibnamefont {Antonius}}, \bibinfo {author}
  {\bibfnamefont {F.}~\bibnamefont {Bruneval}}, \bibinfo {author}
  {\bibfnamefont {F.~H.}\ \bibnamefont {{da Jornada}}}, \bibinfo {author}
  {\bibfnamefont {M.}~\bibnamefont {J.{van Setten}}}, \bibinfo {author}
  {\bibfnamefont {O.~K.}\ \bibnamefont {Orhan}}, \bibinfo {author}
  {\bibfnamefont {D.~D.}\ \bibnamefont {{O'Regan}}}, \bibinfo {author}
  {\bibfnamefont {A.}~\bibnamefont {Canning}}, \bibinfo {author} {\bibfnamefont
  {A.}~\bibnamefont {Ferretti}}, \bibinfo {author} {\bibfnamefont
  {A.}~\bibnamefont {Marini}}, \bibinfo {author} {\bibfnamefont {G.-M.}\
  \bibnamefont {Rignanese}}, \bibinfo {author} {\bibfnamefont {J.}~\bibnamefont
  {Deslippe}}, \bibinfo {author} {\bibfnamefont {S.~G.}\ \bibnamefont
  {Louie}},\ and\ \bibinfo {author} {\bibfnamefont {J.~B.}\ \bibnamefont
  {Neaton}},\ }\href@noop {} {\bibfield  {journal} {\bibinfo  {journal}
  {Comput. Phys. Commun.}\ }\textbf {\bibinfo {volume} {255}},\ \bibinfo
  {pages} {107242} (\bibinfo {year} {2020})}\BibitemShut {NoStop}%
\bibitem [{\citenamefont {Ren}\ \emph {et~al.}(2021)\citenamefont {Ren},
  \citenamefont {Merz}, \citenamefont {Jiang}, \citenamefont {Yao},
  \citenamefont {Rampp}, \citenamefont {Lederer}, \citenamefont {Blum},\ and\
  \citenamefont {Scheffler}}]{Ren/etal:2021}%
  \BibitemOpen
  \bibfield  {author} {\bibinfo {author} {\bibfnamefont {X.}~\bibnamefont
  {Ren}}, \bibinfo {author} {\bibfnamefont {F.}~\bibnamefont {Merz}}, \bibinfo
  {author} {\bibfnamefont {H.}~\bibnamefont {Jiang}}, \bibinfo {author}
  {\bibfnamefont {Y.}~\bibnamefont {Yao}}, \bibinfo {author} {\bibfnamefont
  {M.}~\bibnamefont {Rampp}}, \bibinfo {author} {\bibfnamefont
  {H.}~\bibnamefont {Lederer}}, \bibinfo {author} {\bibfnamefont
  {V.}~\bibnamefont {Blum}},\ and\ \bibinfo {author} {\bibfnamefont
  {M.}~\bibnamefont {Scheffler}},\ }\href@noop {} {\bibfield  {journal}
  {\bibinfo  {journal} {Phys. Rev. Mater.}\ }\textbf {\bibinfo {volume} {5}},\
  \bibinfo {pages} {013807} (\bibinfo {year} {2021})}\BibitemShut {NoStop}%
\bibitem [{\citenamefont {Rostgaard}\ \emph {et~al.}(2010)\citenamefont
  {Rostgaard}, \citenamefont {Jacobsen},\ and\ \citenamefont
  {Thygesen}}]{Rostgaard/Jacobsen/Thygesen:2010}%
  \BibitemOpen
  \bibfield  {author} {\bibinfo {author} {\bibfnamefont {C.}~\bibnamefont
  {Rostgaard}}, \bibinfo {author} {\bibfnamefont {K.~W.}\ \bibnamefont
  {Jacobsen}},\ and\ \bibinfo {author} {\bibfnamefont {K.~S.}\ \bibnamefont
  {Thygesen}},\ }\href@noop {} {\bibfield  {journal} {\bibinfo  {journal}
  {{Phys. Rev. B}}\ }\textbf {\bibinfo {volume} {{81}}} (\bibinfo {year}
  {{2010}})}\BibitemShut {NoStop}%
\bibitem [{\citenamefont {Blase}\ \emph {et~al.}(2011)\citenamefont {Blase},
  \citenamefont {Attaccalite},\ and\ \citenamefont
  {Olevano}}]{Blase/Attaccalite/Olevano:2011}%
  \BibitemOpen
  \bibfield  {author} {\bibinfo {author} {\bibfnamefont {X.}~\bibnamefont
  {Blase}}, \bibinfo {author} {\bibfnamefont {C.}~\bibnamefont {Attaccalite}},\
  and\ \bibinfo {author} {\bibfnamefont {V.}~\bibnamefont {Olevano}},\
  }\href@noop {} {\bibfield  {journal} {\bibinfo  {journal} {Phys. Rev. B}\
  }\textbf {\bibinfo {volume} {83}},\ \bibinfo {pages} {115103} (\bibinfo
  {year} {2011})}\BibitemShut {NoStop}%
\bibitem [{\citenamefont {Foerster}\ \emph {et~al.}(2011)\citenamefont
  {Foerster}, \citenamefont {Koval},\ and\ \citenamefont
  {S{\'a}nchez-Portal}}]{Foerster/etal:2011}%
  \BibitemOpen
  \bibfield  {author} {\bibinfo {author} {\bibfnamefont {D.}~\bibnamefont
  {Foerster}}, \bibinfo {author} {\bibfnamefont {P.}~\bibnamefont {Koval}},\
  and\ \bibinfo {author} {\bibfnamefont {D.}~\bibnamefont
  {S{\'a}nchez-Portal}},\ }\href@noop {} {\bibfield  {journal} {\bibinfo
  {journal} {J. Chem. Phys.}\ }\textbf {\bibinfo {volume} {135}},\ \bibinfo
  {pages} {074105} (\bibinfo {year} {2011})}\BibitemShut {NoStop}%
\bibitem [{\citenamefont {Faber}\ \emph {et~al.}(2012)\citenamefont {Faber},
  \citenamefont {Duchemin}, \citenamefont {Deutsch}, \citenamefont
  {Attaccalite}, \citenamefont {Olevano},\ and\ \citenamefont
  {Blase}}]{Blase_review:2012}%
  \BibitemOpen
  \bibfield  {author} {\bibinfo {author} {\bibfnamefont {C.}~\bibnamefont
  {Faber}}, \bibinfo {author} {\bibfnamefont {I.}~\bibnamefont {Duchemin}},
  \bibinfo {author} {\bibfnamefont {T.}~\bibnamefont {Deutsch}}, \bibinfo
  {author} {\bibfnamefont {C.}~\bibnamefont {Attaccalite}}, \bibinfo {author}
  {\bibfnamefont {V.}~\bibnamefont {Olevano}},\ and\ \bibinfo {author}
  {\bibfnamefont {X.}~\bibnamefont {Blase}},\ }\href
  {https://doi.org/10.1007/s10853-012-6401-7} {\bibfield  {journal} {\bibinfo
  {journal} {J. Mater. Sci.}\ }\textbf {\bibinfo {volume} {47}},\ \bibinfo
  {pages} {7472} (\bibinfo {year} {2012})}\BibitemShut {NoStop}%
\bibitem [{\citenamefont {Ren}\ \emph {et~al.}(2012)\citenamefont {Ren},
  \citenamefont {Rinke}, \citenamefont {Blum}, \citenamefont {Wieferink},
  \citenamefont {Tkatchenko}, \citenamefont {Sanfilippo}, \citenamefont
  {Reuter},\ and\ \citenamefont {Scheffler}}]{Ren/etal:2012}%
  \BibitemOpen
  \bibfield  {author} {\bibinfo {author} {\bibfnamefont {X.}~\bibnamefont
  {Ren}}, \bibinfo {author} {\bibfnamefont {P.}~\bibnamefont {Rinke}}, \bibinfo
  {author} {\bibfnamefont {V.}~\bibnamefont {Blum}}, \bibinfo {author}
  {\bibfnamefont {J.}~\bibnamefont {Wieferink}}, \bibinfo {author}
  {\bibfnamefont {A.}~\bibnamefont {Tkatchenko}}, \bibinfo {author}
  {\bibfnamefont {A.}~\bibnamefont {Sanfilippo}}, \bibinfo {author}
  {\bibfnamefont {K.}~\bibnamefont {Reuter}},\ and\ \bibinfo {author}
  {\bibfnamefont {M.}~\bibnamefont {Scheffler}},\ }\href
  {https://doi.org/10.1088/1367-2630/14/5/053020} {\bibfield  {journal}
  {\bibinfo  {journal} {New J. Phys.}\ }\textbf {\bibinfo {volume} {14}},\
  \bibinfo {pages} {053020} (\bibinfo {year} {2012})}\BibitemShut {NoStop}%
\bibitem [{\citenamefont {Caruso}\ \emph {et~al.}(2012)\citenamefont {Caruso},
  \citenamefont {Rinke}, \citenamefont {Ren}, \citenamefont {Scheffler},\ and\
  \citenamefont {Rubio}}]{Caruso/etal:2012}%
  \BibitemOpen
  \bibfield  {author} {\bibinfo {author} {\bibfnamefont {F.}~\bibnamefont
  {Caruso}}, \bibinfo {author} {\bibfnamefont {P.}~\bibnamefont {Rinke}},
  \bibinfo {author} {\bibfnamefont {X.}~\bibnamefont {Ren}}, \bibinfo {author}
  {\bibfnamefont {M.}~\bibnamefont {Scheffler}},\ and\ \bibinfo {author}
  {\bibfnamefont {A.}~\bibnamefont {Rubio}},\ }\href
  {https://link.aps.org/doi/10.1103/PhysRevB.86.081102} {\bibfield  {journal}
  {\bibinfo  {journal} {Phys. Rev. B}\ }\textbf {\bibinfo {volume} {86}},\
  \bibinfo {pages} {081102(R)} (\bibinfo {year} {2012})}\BibitemShut {NoStop}%
\bibitem [{\citenamefont {Bruneval}\ and\ \citenamefont
  {Marques}(2013)}]{Bruneval/Marques:2013}%
  \BibitemOpen
  \bibfield  {author} {\bibinfo {author} {\bibfnamefont {F.}~\bibnamefont
  {Bruneval}}\ and\ \bibinfo {author} {\bibfnamefont {M.~A.~L.}\ \bibnamefont
  {Marques}},\ }\href@noop {} {\bibfield  {journal} {\bibinfo  {journal} {J.
  Chem. Theo. Comp.}\ }\textbf {\bibinfo {volume} {9}},\ \bibinfo {pages} {324}
  (\bibinfo {year} {2013})}\BibitemShut {NoStop}%
\bibitem [{\citenamefont {{van Setten}}\ \emph {et~al.}(2013)\citenamefont
  {{van Setten}}, \citenamefont {Weigend},\ and\ \citenamefont
  {Evers}}]{Setten/etal:2013}%
  \BibitemOpen
  \bibfield  {author} {\bibinfo {author} {\bibfnamefont {M.~J.}\ \bibnamefont
  {{van Setten}}}, \bibinfo {author} {\bibfnamefont {F.}~\bibnamefont
  {Weigend}},\ and\ \bibinfo {author} {\bibfnamefont {F.}~\bibnamefont
  {Evers}},\ }\href@noop {} {\bibfield  {journal} {\bibinfo  {journal} {J.
  Chem. Theo. Comp.}\ }\textbf {\bibinfo {volume} {9}},\ \bibinfo {pages} {232}
  (\bibinfo {year} {2013})}\BibitemShut {NoStop}%
\bibitem [{\citenamefont {van Setten}\ \emph {et~al.}(2015)\citenamefont {van
  Setten}, \citenamefont {Caruso}, \citenamefont {Sharifzadeh}, \citenamefont
  {Ren}, \citenamefont {Scheffler}, \citenamefont {Liu}, \citenamefont
  {Lischner}, \citenamefont {Lin}, \citenamefont {Deslippe}, \citenamefont
  {Louie}, \citenamefont {Yang}, \citenamefont {Weigend}, \citenamefont
  {Neaton}, \citenamefont {Evers},\ and\ \citenamefont
  {Rinke}}]{Setten/etal:2015}%
  \BibitemOpen
  \bibfield  {author} {\bibinfo {author} {\bibfnamefont {M.~J.}\ \bibnamefont
  {van Setten}}, \bibinfo {author} {\bibfnamefont {F.}~\bibnamefont {Caruso}},
  \bibinfo {author} {\bibfnamefont {S.}~\bibnamefont {Sharifzadeh}}, \bibinfo
  {author} {\bibfnamefont {X.}~\bibnamefont {Ren}}, \bibinfo {author}
  {\bibfnamefont {M.}~\bibnamefont {Scheffler}}, \bibinfo {author}
  {\bibfnamefont {F.}~\bibnamefont {Liu}}, \bibinfo {author} {\bibfnamefont
  {J.}~\bibnamefont {Lischner}}, \bibinfo {author} {\bibfnamefont
  {L.}~\bibnamefont {Lin}}, \bibinfo {author} {\bibfnamefont {J.~R.}\
  \bibnamefont {Deslippe}}, \bibinfo {author} {\bibfnamefont {S.~G.}\
  \bibnamefont {Louie}}, \bibinfo {author} {\bibfnamefont {C.}~\bibnamefont
  {Yang}}, \bibinfo {author} {\bibfnamefont {F.}~\bibnamefont {Weigend}},
  \bibinfo {author} {\bibfnamefont {J.~B.}\ \bibnamefont {Neaton}}, \bibinfo
  {author} {\bibfnamefont {F.}~\bibnamefont {Evers}},\ and\ \bibinfo {author}
  {\bibfnamefont {P.}~\bibnamefont {Rinke}},\ }\href
  {https://doi.org/10.1021/acs.jctc.5b00453} {\bibfield  {journal} {\bibinfo
  {journal} {J. Chem. Theory Comput.}\ }\textbf {\bibinfo {volume} {11}},\
  \bibinfo {pages} {5665} (\bibinfo {year} {2015})}\BibitemShut {NoStop}%
\bibitem [{\citenamefont {Knight}\ \emph {et~al.}(2016)\citenamefont {Knight},
  \citenamefont {Wang}, \citenamefont {Gallandi}, \citenamefont
  {Dolgounitcheva}, \citenamefont {Ren}, \citenamefont {Ortiz}, \citenamefont
  {Rinke}, \citenamefont {Körzdörfer},\ and\ \citenamefont
  {Marom}}]{Knight/etal:2016}%
  \BibitemOpen
  \bibfield  {author} {\bibinfo {author} {\bibfnamefont {J.~W.}\ \bibnamefont
  {Knight}}, \bibinfo {author} {\bibfnamefont {X.}~\bibnamefont {Wang}},
  \bibinfo {author} {\bibfnamefont {L.}~\bibnamefont {Gallandi}}, \bibinfo
  {author} {\bibfnamefont {O.}~\bibnamefont {Dolgounitcheva}}, \bibinfo
  {author} {\bibfnamefont {X.}~\bibnamefont {Ren}}, \bibinfo {author}
  {\bibfnamefont {J.~V.}\ \bibnamefont {Ortiz}}, \bibinfo {author}
  {\bibfnamefont {P.}~\bibnamefont {Rinke}}, \bibinfo {author} {\bibfnamefont
  {T.}~\bibnamefont {Körzdörfer}},\ and\ \bibinfo {author} {\bibfnamefont
  {N.}~\bibnamefont {Marom}},\ }\href
  {https://doi.org/10.1021/acs.jctc.5b00871} {\bibfield  {journal} {\bibinfo
  {journal} {J. Chem. Theory Comput.}\ }\textbf {\bibinfo {volume} {12}},\
  \bibinfo {pages} {615} (\bibinfo {year} {2016})}\BibitemShut {NoStop}%
\bibitem [{\citenamefont {Maggio}\ and\ \citenamefont
  {Kresse}(2017)}]{Maggio/Kresse:2017}%
  \BibitemOpen
  \bibfield  {author} {\bibinfo {author} {\bibfnamefont {E.}~\bibnamefont
  {Maggio}}\ and\ \bibinfo {author} {\bibfnamefont {G.}~\bibnamefont
  {Kresse}},\ }\href@noop {} {\bibfield  {journal} {\bibinfo  {journal} {J.
  Chem. Theory Comput.}\ }\textbf {\bibinfo {volume} {13}},\ \bibinfo {pages}
  {4765} (\bibinfo {year} {2017})}\BibitemShut {NoStop}%
\bibitem [{\citenamefont {Lange}\ and\ \citenamefont
  {Berkelbach}(2018)}]{Lange/Berkelbach:2018}%
  \BibitemOpen
  \bibfield  {author} {\bibinfo {author} {\bibfnamefont {M.}~\bibnamefont
  {Lange}}\ and\ \bibinfo {author} {\bibfnamefont {T.~C.}\ \bibnamefont
  {Berkelbach}},\ }\href {https://doi.org/10.1021/acs.jctc.8b00455} {\bibfield
  {journal} {\bibinfo  {journal} {J. Chem. Theory Comput.}\ }\textbf {\bibinfo
  {volume} {14}},\ \bibinfo {pages} {4224} (\bibinfo {year}
  {2018})}\BibitemShut {NoStop}%
\bibitem [{\citenamefont {Golze}\ \emph {et~al.}(2018)\citenamefont {Golze},
  \citenamefont {Wilhelm}, \citenamefont {{van Setten}},\ and\ \citenamefont
  {Rinke}}]{Golze/etal:2018}%
  \BibitemOpen
  \bibfield  {author} {\bibinfo {author} {\bibfnamefont {D.}~\bibnamefont
  {Golze}}, \bibinfo {author} {\bibfnamefont {J.}~\bibnamefont {Wilhelm}},
  \bibinfo {author} {\bibfnamefont {M.~J.}\ \bibnamefont {{van Setten}}},\ and\
  \bibinfo {author} {\bibfnamefont {P.}~\bibnamefont {Rinke}},\ }\href@noop {}
  {\bibfield  {journal} {\bibinfo  {journal} {J. Chem. Theory Comput.}\
  }\textbf {\bibinfo {volume} {14}},\ \bibinfo {pages} {4856} (\bibinfo {year}
  {2018})}\BibitemShut {NoStop}%
\bibitem [{\citenamefont {Bakhsh}\ \emph {et~al.}(2021)\citenamefont {Bakhsh},
  \citenamefont {Liu}, \citenamefont {Wang}, \citenamefont {He},\ and\
  \citenamefont {Ren}}]{Sunila/etal:2021}%
  \BibitemOpen
  \bibfield  {author} {\bibinfo {author} {\bibfnamefont {S.}~\bibnamefont
  {Bakhsh}}, \bibinfo {author} {\bibfnamefont {X.}~\bibnamefont {Liu}},
  \bibinfo {author} {\bibfnamefont {Y.}~\bibnamefont {Wang}}, \bibinfo {author}
  {\bibfnamefont {L.}~\bibnamefont {He}},\ and\ \bibinfo {author}
  {\bibfnamefont {X.}~\bibnamefont {Ren}},\ }\href
  {https://doi.org/10.1021/acs.jpca.0c08960} {\bibfield  {journal} {\bibinfo
  {journal} {The Journal of Physical Chemistry A}\ }\textbf {\bibinfo {volume}
  {125}},\ \bibinfo {pages} {1424} (\bibinfo {year} {2021})},\ \bibinfo {note}
  {pMID: 33591198},\ \Eprint
  {https://arxiv.org/abs/https://doi.org/10.1021/acs.jpca.0c08960}
  {https://doi.org/10.1021/acs.jpca.0c08960} \BibitemShut {NoStop}%
\bibitem [{\citenamefont {Hybertsen}\ and\ \citenamefont
  {Louie}(1985)}]{Hybertsen/Louie:1985}%
  \BibitemOpen
  \bibfield  {author} {\bibinfo {author} {\bibfnamefont {M.~S.}\ \bibnamefont
  {Hybertsen}}\ and\ \bibinfo {author} {\bibfnamefont {S.~G.}\ \bibnamefont
  {Louie}},\ }\href {https://doi.org/10.1103/PhysRevLett.55.1418} {\bibfield
  {journal} {\bibinfo  {journal} {Phys. Rev. Lett.}\ }\textbf {\bibinfo
  {volume} {55}},\ \bibinfo {pages} {1418} (\bibinfo {year}
  {1985})}\BibitemShut {NoStop}%
\bibitem [{\citenamefont {von Barth}\ and\ \citenamefont
  {Holm}(1996)}]{vonBarth/Holm:1996}%
  \BibitemOpen
  \bibfield  {author} {\bibinfo {author} {\bibfnamefont {U.}~\bibnamefont {von
  Barth}}\ and\ \bibinfo {author} {\bibfnamefont {B.}~\bibnamefont {Holm}},\
  }\href@noop {} {\bibfield  {journal} {\bibinfo  {journal} {Phys.\ Rev. B}\
  }\textbf {\bibinfo {volume} {54}},\ \bibinfo {pages} {8411} (\bibinfo {year}
  {1996})}\BibitemShut {NoStop}%
\bibitem [{\citenamefont {Faleev}\ \emph {et~al.}(2004)\citenamefont {Faleev},
  \citenamefont {\mbox{van Schilfgaarde}},\ and\ \citenamefont
  {Kotani}}]{Faleev/Schilfgaarde/Kotani:2004}%
  \BibitemOpen
  \bibfield  {author} {\bibinfo {author} {\bibfnamefont {S.}~\bibnamefont
  {Faleev}}, \bibinfo {author} {\bibfnamefont {M.}~\bibnamefont {\mbox{van
  Schilfgaarde}}},\ and\ \bibinfo {author} {\bibfnamefont {T.}~\bibnamefont
  {Kotani}},\ }\href@noop {} {\bibfield  {journal} {\bibinfo  {journal} {Phys.\
  Rev.\ Lett.}\ }\textbf {\bibinfo {volume} {93}},\ \bibinfo {pages} {126406}
  (\bibinfo {year} {2004})}\BibitemShut {NoStop}%
\bibitem [{\citenamefont {Ku}\ and\ \citenamefont
  {Eguiluz}(2002)}]{Ku/Eguiluz:2002}%
  \BibitemOpen
  \bibfield  {author} {\bibinfo {author} {\bibfnamefont {W.}~\bibnamefont
  {Ku}}\ and\ \bibinfo {author} {\bibfnamefont {A.~G.}\ \bibnamefont
  {Eguiluz}},\ }\href@noop {} {\bibfield  {journal} {\bibinfo  {journal} {Phys.
  Rev. Lett.}\ }\textbf {\bibinfo {volume} {89}},\ \bibinfo {pages} {126401}
  (\bibinfo {year} {2002})}\BibitemShut {NoStop}%
\bibitem [{\citenamefont {Stan}\ \emph {et~al.}(2006)\citenamefont {Stan},
  \citenamefont {Dahlen},\ and\ \citenamefont {{van
  Leeuwen}}}]{Stan/Dahlen/Leeuwen:2006}%
  \BibitemOpen
  \bibfield  {author} {\bibinfo {author} {\bibfnamefont {A.}~\bibnamefont
  {Stan}}, \bibinfo {author} {\bibfnamefont {N.~E.}\ \bibnamefont {Dahlen}},\
  and\ \bibinfo {author} {\bibfnamefont {R.}~\bibnamefont {{van Leeuwen}}},\
  }\href@noop {} {\bibfield  {journal} {\bibinfo  {journal} {Europhys.\ Lett.}\
  }\textbf {\bibinfo {volume} {76}},\ \bibinfo {pages} {298} (\bibinfo {year}
  {2006})}\BibitemShut {NoStop}%
\bibitem [{\citenamefont {Caruso}\ \emph {et~al.}(2013)\citenamefont {Caruso},
  \citenamefont {Rinke}, \citenamefont {Ren}, \citenamefont {Scheffler},\ and\
  \citenamefont {Rubio}}]{Caruso/etal:2013}%
  \BibitemOpen
  \bibfield  {author} {\bibinfo {author} {\bibfnamefont {F.}~\bibnamefont
  {Caruso}}, \bibinfo {author} {\bibfnamefont {P.}~\bibnamefont {Rinke}},
  \bibinfo {author} {\bibfnamefont {X.}~\bibnamefont {Ren}}, \bibinfo {author}
  {\bibfnamefont {M.}~\bibnamefont {Scheffler}},\ and\ \bibinfo {author}
  {\bibfnamefont {A.}~\bibnamefont {Rubio}},\ }\href
  {https://link.aps.org/doi/10.1103/PhysRevB.88.075105} {\bibfield  {journal}
  {\bibinfo  {journal} {Phys. Rev. B}\ }\textbf {\bibinfo {volume} {88}},\
  \bibinfo {pages} {075105} (\bibinfo {year} {2013})}\BibitemShut {NoStop}%
\bibitem [{\citenamefont {Kutepov}(2016)}]{Kutepov:2016}%
  \BibitemOpen
  \bibfield  {author} {\bibinfo {author} {\bibfnamefont {A.~L.}\ \bibnamefont
  {Kutepov}},\ }\href {https://doi.org/10.1103/PhysRevB.94.155101} {\bibfield
  {journal} {\bibinfo  {journal} {Phys. Rev. B}\ }\textbf {\bibinfo {volume}
  {94}},\ \bibinfo {pages} {155101} (\bibinfo {year} {2016})}\BibitemShut
  {NoStop}%
\bibitem [{\citenamefont {Kutepov}(2017)}]{Kutepov:2017}%
  \BibitemOpen
  \bibfield  {author} {\bibinfo {author} {\bibfnamefont {A.~L.}\ \bibnamefont
  {Kutepov}},\ }\href {https://doi.org/10.1103/PhysRevB.95.195120} {\bibfield
  {journal} {\bibinfo  {journal} {Phys. Rev. B}\ }\textbf {\bibinfo {volume}
  {95}},\ \bibinfo {pages} {195120} (\bibinfo {year} {2017})}\BibitemShut
  {NoStop}%
\bibitem [{\citenamefont {Fuchs}\ \emph {et~al.}(2007)\citenamefont {Fuchs},
  \citenamefont {Furthm\"uller}, \citenamefont {Bechstedt}, \citenamefont
  {Shishkin},\ and\ \citenamefont {Kresse}}]{Fuchs/etal:2007}%
  \BibitemOpen
  \bibfield  {author} {\bibinfo {author} {\bibfnamefont {F.}~\bibnamefont
  {Fuchs}}, \bibinfo {author} {\bibfnamefont {J.}~\bibnamefont
  {Furthm\"uller}}, \bibinfo {author} {\bibfnamefont {F.}~\bibnamefont
  {Bechstedt}}, \bibinfo {author} {\bibfnamefont {M.}~\bibnamefont
  {Shishkin}},\ and\ \bibinfo {author} {\bibfnamefont {G.}~\bibnamefont
  {Kresse}},\ }\href@noop {} {\bibfield  {journal} {\bibinfo  {journal} {Phys.
  Rev. B}\ }\textbf {\bibinfo {volume} {76}},\ \bibinfo {pages} {115109}
  (\bibinfo {year} {2007})}\BibitemShut {NoStop}%
\bibitem [{\citenamefont {Jiang}\ \emph {et~al.}(2009)\citenamefont {Jiang},
  \citenamefont {G{\'o}mez-Abal}, \citenamefont {Rinke},\ and\ \citenamefont
  {Scheffler}}]{Jiang/etal:2009}%
  \BibitemOpen
  \bibfield  {author} {\bibinfo {author} {\bibfnamefont {H.}~\bibnamefont
  {Jiang}}, \bibinfo {author} {\bibfnamefont {R.~I.}\ \bibnamefont
  {G{\'o}mez-Abal}}, \bibinfo {author} {\bibfnamefont {P.}~\bibnamefont
  {Rinke}},\ and\ \bibinfo {author} {\bibfnamefont {M.}~\bibnamefont
  {Scheffler}},\ }\href@noop {} {\bibfield  {journal} {\bibinfo  {journal}
  {Phys. Rev. Lett.}\ }\textbf {\bibinfo {volume} {102}},\ \bibinfo {pages}
  {126403} (\bibinfo {year} {2009})}\BibitemShut {NoStop}%
\bibitem [{\citenamefont {Marom}\ \emph {et~al.}(2012)\citenamefont {Marom},
  \citenamefont {Caruso}, \citenamefont {Ren}, \citenamefont {Hofmann},
  \citenamefont {K{\"o}rzd{\"o}rfer}, \citenamefont {Chelikowsky},
  \citenamefont {Rubio}, \citenamefont {Scheffler},\ and\ \citenamefont
  {Rinke}}]{Marom/etal:2012}%
  \BibitemOpen
  \bibfield  {author} {\bibinfo {author} {\bibfnamefont {N.}~\bibnamefont
  {Marom}}, \bibinfo {author} {\bibfnamefont {F.}~\bibnamefont {Caruso}},
  \bibinfo {author} {\bibfnamefont {X.}~\bibnamefont {Ren}}, \bibinfo {author}
  {\bibfnamefont {O.~T.}\ \bibnamefont {Hofmann}}, \bibinfo {author}
  {\bibfnamefont {T.}~\bibnamefont {K{\"o}rzd{\"o}rfer}}, \bibinfo {author}
  {\bibfnamefont {J.~R.}\ \bibnamefont {Chelikowsky}}, \bibinfo {author}
  {\bibfnamefont {A.}~\bibnamefont {Rubio}}, \bibinfo {author} {\bibfnamefont
  {M.}~\bibnamefont {Scheffler}},\ and\ \bibinfo {author} {\bibfnamefont
  {P.}~\bibnamefont {Rinke}},\ }\href@noop {} {\bibfield  {journal} {\bibinfo
  {journal} {Phys. Rev. B}\ }\textbf {\bibinfo {volume} {86}},\ \bibinfo
  {pages} {245127} (\bibinfo {year} {2012})}\BibitemShut {NoStop}%
\bibitem [{\citenamefont {Ren}\ \emph {et~al.}(2015)\citenamefont {Ren},
  \citenamefont {Marom}, \citenamefont {Caruso}, \citenamefont {Scheffler},\
  and\ \citenamefont {Rinke}}]{Ren/etal:2015}%
  \BibitemOpen
  \bibfield  {author} {\bibinfo {author} {\bibfnamefont {X.}~\bibnamefont
  {Ren}}, \bibinfo {author} {\bibfnamefont {N.}~\bibnamefont {Marom}}, \bibinfo
  {author} {\bibfnamefont {F.}~\bibnamefont {Caruso}}, \bibinfo {author}
  {\bibfnamefont {M.}~\bibnamefont {Scheffler}},\ and\ \bibinfo {author}
  {\bibfnamefont {P.}~\bibnamefont {Rinke}},\ }\href
  {https://doi.org/10.1103/PhysRevB.92.081104} {\bibfield  {journal} {\bibinfo
  {journal} {Phys. Rev. B}\ }\textbf {\bibinfo {volume} {92}},\ \bibinfo
  {pages} {081104} (\bibinfo {year} {2015})}\BibitemShut {NoStop}%
\bibitem [{\citenamefont {Dvorak}\ and\ \citenamefont
  {Rinke}(2019)}]{Dvorak/Rinke:2019}%
  \BibitemOpen
  \bibfield  {author} {\bibinfo {author} {\bibfnamefont {M.}~\bibnamefont
  {Dvorak}}\ and\ \bibinfo {author} {\bibfnamefont {P.}~\bibnamefont {Rinke}},\
  }\href {https://doi.org/10.1103/PhysRevB.99.115134} {\bibfield  {journal}
  {\bibinfo  {journal} {Phys. Rev. B}\ }\textbf {\bibinfo {volume} {99}},\
  \bibinfo {pages} {115134} (\bibinfo {year} {2019})}\BibitemShut {NoStop}%
\bibitem [{\citenamefont {Dvorak}\ \emph {et~al.}(2019)\citenamefont {Dvorak},
  \citenamefont {Golze},\ and\ \citenamefont
  {Rinke}}]{Dvorak/Golze/Rinke:2019}%
  \BibitemOpen
  \bibfield  {author} {\bibinfo {author} {\bibfnamefont {M.}~\bibnamefont
  {Dvorak}}, \bibinfo {author} {\bibfnamefont {D.}~\bibnamefont {Golze}},\ and\
  \bibinfo {author} {\bibfnamefont {P.}~\bibnamefont {Rinke}},\ }\href
  {https://doi.org/10.1103/PhysRevMaterials.3.070801} {\bibfield  {journal}
  {\bibinfo  {journal} {Phys. Rev. Materials}\ }\textbf {\bibinfo {volume}
  {3}},\ \bibinfo {pages} {070801} (\bibinfo {year} {2019})}\BibitemShut
  {NoStop}%
\bibitem [{\citenamefont {Gr\"{u}neis}\ \emph {et~al.}(2014)\citenamefont
  {Gr\"{u}neis}, \citenamefont {Kresse}, \citenamefont {Hinuma},\ and\
  \citenamefont {Oba}}]{Grueneis/etal:2014}%
  \BibitemOpen
  \bibfield  {author} {\bibinfo {author} {\bibfnamefont {A.}~\bibnamefont
  {Gr\"{u}neis}}, \bibinfo {author} {\bibfnamefont {G.}~\bibnamefont {Kresse}},
  \bibinfo {author} {\bibfnamefont {Y.}~\bibnamefont {Hinuma}},\ and\ \bibinfo
  {author} {\bibfnamefont {F.}~\bibnamefont {Oba}},\ }\href@noop {} {\bibfield
  {journal} {\bibinfo  {journal} {Phys. Rev. Lett.}\ }\textbf {\bibinfo
  {volume} {112}},\ \bibinfo {pages} {096401} (\bibinfo {year}
  {2014})}\BibitemShut {NoStop}%
\bibitem [{\citenamefont {{Del Sole}}\ \emph {et~al.}(1994)\citenamefont {{Del
  Sole}}, \citenamefont {Reining},\ and\ \citenamefont
  {Godby}}]{DelSole/etal:1994}%
  \BibitemOpen
  \bibfield  {author} {\bibinfo {author} {\bibfnamefont {R.}~\bibnamefont {{Del
  Sole}}}, \bibinfo {author} {\bibfnamefont {L.}~\bibnamefont {Reining}},\ and\
  \bibinfo {author} {\bibfnamefont {R.~W.}\ \bibnamefont {Godby}},\ }\href@noop
  {} {\bibfield  {journal} {\bibinfo  {journal} {Phys. Rev. B}\ }\textbf
  {\bibinfo {volume} {49}},\ \bibinfo {pages} {8024} (\bibinfo {year}
  {1994})}\BibitemShut {NoStop}%
\bibitem [{\citenamefont {Shirley}(1996)}]{Shirley:1996}%
  \BibitemOpen
  \bibfield  {author} {\bibinfo {author} {\bibfnamefont {E.~L.}\ \bibnamefont
  {Shirley}},\ }\href@noop {} {\bibfield  {journal} {\bibinfo  {journal} {Phys.
  Rev. B}\ }\textbf {\bibinfo {volume} {54}},\ \bibinfo {pages} {7758}
  (\bibinfo {year} {1996})}\BibitemShut {NoStop}%
\bibitem [{\citenamefont {Bobbert}\ and\ \citenamefont {{van
  Haeringen}}(1994)}]{Bobbert/Haeringen:1994}%
  \BibitemOpen
  \bibfield  {author} {\bibinfo {author} {\bibfnamefont {P.~A.}\ \bibnamefont
  {Bobbert}}\ and\ \bibinfo {author} {\bibfnamefont {W.}~\bibnamefont {{van
  Haeringen}}},\ }\href@noop {} {\bibfield  {journal} {\bibinfo  {journal}
  {Phys. Rev. B}\ }\textbf {\bibinfo {volume} {49}},\ \bibinfo {pages} {10326}
  (\bibinfo {year} {1994})}\BibitemShut {NoStop}%
\bibitem [{\citenamefont {Bruneval}\ \emph {et~al.}(2005)\citenamefont
  {Bruneval}, \citenamefont {Sottile}, \citenamefont {Olevano}, \citenamefont
  {{Del Sole}},\ and\ \citenamefont {Reining}}]{Bruneval/etal:2005}%
  \BibitemOpen
  \bibfield  {author} {\bibinfo {author} {\bibfnamefont {F.}~\bibnamefont
  {Bruneval}}, \bibinfo {author} {\bibfnamefont {F.}~\bibnamefont {Sottile}},
  \bibinfo {author} {\bibfnamefont {V.}~\bibnamefont {Olevano}}, \bibinfo
  {author} {\bibfnamefont {R.}~\bibnamefont {{Del Sole}}},\ and\ \bibinfo
  {author} {\bibfnamefont {L.}~\bibnamefont {Reining}},\ }\href@noop {}
  {\bibfield  {journal} {\bibinfo  {journal} {Phys. Rev. Lett.}\ }\textbf
  {\bibinfo {volume} {94}},\ \bibinfo {pages} {186402} (\bibinfo {year}
  {2005})}\BibitemShut {NoStop}%
\bibitem [{\citenamefont {Schindlmayr}\ and\ \citenamefont
  {Godby}(1998)}]{Schindlmayr/Godby:1998}%
  \BibitemOpen
  \bibfield  {author} {\bibinfo {author} {\bibfnamefont {A.}~\bibnamefont
  {Schindlmayr}}\ and\ \bibinfo {author} {\bibfnamefont {R.~W.}\ \bibnamefont
  {Godby}},\ }\href@noop {} {\bibfield  {journal} {\bibinfo  {journal} {Phys.
  Rev. Lett.}\ }\textbf {\bibinfo {volume} {80}},\ \bibinfo {pages} {1702}
  (\bibinfo {year} {1998})}\BibitemShut {NoStop}%
\bibitem [{\citenamefont {Maebashi}\ and\ \citenamefont
  {Takada}(2011)}]{Maebashi/Takada:2011}%
  \BibitemOpen
  \bibfield  {author} {\bibinfo {author} {\bibfnamefont {H.}~\bibnamefont
  {Maebashi}}\ and\ \bibinfo {author} {\bibfnamefont {Y.}~\bibnamefont
  {Takada}},\ }\href {https://doi.org/10.1103/PhysRevB.84.245134} {\bibfield
  {journal} {\bibinfo  {journal} {Phys. Rev. B}\ }\textbf {\bibinfo {volume}
  {84}},\ \bibinfo {pages} {245134} (\bibinfo {year} {2011})}\BibitemShut
  {NoStop}%
\bibitem [{\citenamefont {Romaniello}\ \emph {et~al.}(2012)\citenamefont
  {Romaniello}, \citenamefont {Bechstedt},\ and\ \citenamefont
  {Reining}}]{Romaniello/etal:2012}%
  \BibitemOpen
  \bibfield  {author} {\bibinfo {author} {\bibfnamefont {P.}~\bibnamefont
  {Romaniello}}, \bibinfo {author} {\bibfnamefont {F.}~\bibnamefont
  {Bechstedt}},\ and\ \bibinfo {author} {\bibfnamefont {L.}~\bibnamefont
  {Reining}},\ }\href {https://doi.org/10.1103/PhysRevB.85.155131} {\bibfield
  {journal} {\bibinfo  {journal} {Phys. Rev. B}\ }\textbf {\bibinfo {volume}
  {85}},\ \bibinfo {pages} {155131} (\bibinfo {year} {2012})}\BibitemShut
  {NoStop}%
\bibitem [{\citenamefont {Kuwahara}\ \emph {et~al.}(2016)\citenamefont
  {Kuwahara}, \citenamefont {Noguchi},\ and\ \citenamefont
  {Ohno}}]{Kuwahara2016/PhysRevB.94.121116}%
  \BibitemOpen
  \bibfield  {author} {\bibinfo {author} {\bibfnamefont {R.}~\bibnamefont
  {Kuwahara}}, \bibinfo {author} {\bibfnamefont {Y.}~\bibnamefont {Noguchi}},\
  and\ \bibinfo {author} {\bibfnamefont {K.}~\bibnamefont {Ohno}},\ }\href
  {https://doi.org/10.1103/PhysRevB.94.121116} {\bibfield  {journal} {\bibinfo
  {journal} {Phys. Rev. B}\ }\textbf {\bibinfo {volume} {94}},\ \bibinfo
  {pages} {121116} (\bibinfo {year} {2016})}\BibitemShut {NoStop}%
\bibitem [{\citenamefont {Hellgren}\ \emph {et~al.}(2018)\citenamefont
  {Hellgren}, \citenamefont {Colonna},\ and\ \citenamefont
  {de~Gironcoli}}]{Hellgren2018/PhysRevB.98.045117}%
  \BibitemOpen
  \bibfield  {author} {\bibinfo {author} {\bibfnamefont {M.}~\bibnamefont
  {Hellgren}}, \bibinfo {author} {\bibfnamefont {N.}~\bibnamefont {Colonna}},\
  and\ \bibinfo {author} {\bibfnamefont {S.}~\bibnamefont {de~Gironcoli}},\
  }\href {https://doi.org/10.1103/PhysRevB.98.045117} {\bibfield  {journal}
  {\bibinfo  {journal} {Phys. Rev. B}\ }\textbf {\bibinfo {volume} {98}},\
  \bibinfo {pages} {045117} (\bibinfo {year} {2018})}\BibitemShut {NoStop}%
\bibitem [{\citenamefont {Vlček}(2019)}]{Vojtech2019/acs.jctc.9b00317}%
  \BibitemOpen
  \bibfield  {author} {\bibinfo {author} {\bibfnamefont {V.}~\bibnamefont
  {Vlček}},\ }\href {https://doi.org/10.1021/acs.jctc.9b00317} {\bibfield
  {journal} {\bibinfo  {journal} {J. Chem Theory and Comput.}\ }\textbf
  {\bibinfo {volume} {15}},\ \bibinfo {pages} {6254} (\bibinfo {year}
  {2019})}\BibitemShut {NoStop}%
\bibitem [{\citenamefont {Pavlyukh}\ \emph
  {et~al.}(2020{\natexlab{a}})\citenamefont {Pavlyukh}, \citenamefont
  {Stefanucci},\ and\ \citenamefont {van
  Leeuwen}}]{Leeuwen2020/PhysRevB.102.045121}%
  \BibitemOpen
  \bibfield  {author} {\bibinfo {author} {\bibfnamefont {Y.}~\bibnamefont
  {Pavlyukh}}, \bibinfo {author} {\bibfnamefont {G.}~\bibnamefont
  {Stefanucci}},\ and\ \bibinfo {author} {\bibfnamefont {R.}~\bibnamefont {van
  Leeuwen}},\ }\href {https://doi.org/10.1103/PhysRevB.102.045121} {\bibfield
  {journal} {\bibinfo  {journal} {Phys. Rev. B}\ }\textbf {\bibinfo {volume}
  {102}},\ \bibinfo {pages} {045121} (\bibinfo {year}
  {2020}{\natexlab{a}})}\BibitemShut {NoStop}%
\bibitem [{\citenamefont {Freeman}(1977)}]{Freeman:1977}%
  \BibitemOpen
  \bibfield  {author} {\bibinfo {author} {\bibfnamefont {D.~L.}\ \bibnamefont
  {Freeman}},\ }\href@noop {} {\bibfield  {journal} {\bibinfo  {journal} {Phys.
  Rev. B}\ }\textbf {\bibinfo {volume} {15}},\ \bibinfo {pages} {5512}
  (\bibinfo {year} {1977})}\BibitemShut {NoStop}%
\bibitem [{\citenamefont {Gr{\"u}neis}\ \emph {et~al.}(2009)\citenamefont
  {Gr{\"u}neis}, \citenamefont {Marsman}, \citenamefont {Harl}, \citenamefont
  {Schimka},\ and\ \citenamefont {Kresse}}]{Grueneis/etal:2009}%
  \BibitemOpen
  \bibfield  {author} {\bibinfo {author} {\bibfnamefont {A.}~\bibnamefont
  {Gr{\"u}neis}}, \bibinfo {author} {\bibfnamefont {M.}~\bibnamefont
  {Marsman}}, \bibinfo {author} {\bibfnamefont {J.}~\bibnamefont {Harl}},
  \bibinfo {author} {\bibfnamefont {L.}~\bibnamefont {Schimka}},\ and\ \bibinfo
  {author} {\bibfnamefont {G.}~\bibnamefont {Kresse}},\ }\href@noop {}
  {\bibfield  {journal} {\bibinfo  {journal} {J. Chem. Phys.}\ }\textbf
  {\bibinfo {volume} {131}},\ \bibinfo {pages} {154115} (\bibinfo {year}
  {2009})}\BibitemShut {NoStop}%
\bibitem [{\citenamefont {Jansen}\ \emph {et~al.}(2010)\citenamefont {Jansen},
  \citenamefont {Liu},\ and\ \citenamefont
  {{\'A}ngy{\'a}n}}]{Jansen/etal:2010}%
  \BibitemOpen
  \bibfield  {author} {\bibinfo {author} {\bibfnamefont {G.}~\bibnamefont
  {Jansen}}, \bibinfo {author} {\bibfnamefont {R.-F.}\ \bibnamefont {Liu}},\
  and\ \bibinfo {author} {\bibfnamefont {J.~G.}\ \bibnamefont
  {{\'A}ngy{\'a}n}},\ }\href@noop {} {\bibfield  {journal} {\bibinfo  {journal}
  {J. Chem. Phys.}\ }\textbf {\bibinfo {volume} {133}},\ \bibinfo {pages}
  {154106} (\bibinfo {year} {2010})}\BibitemShut {NoStop}%
\bibitem [{\citenamefont {Paier}\ \emph {et~al.}(2012)\citenamefont {Paier},
  \citenamefont {Ren}, \citenamefont {Rinke}, \citenamefont {Scuseria},
  \citenamefont {Gr{\"u}neis}, \citenamefont {Kresse},\ and\ \citenamefont
  {Scheffler}}]{Paier/etal:2012}%
  \BibitemOpen
  \bibfield  {author} {\bibinfo {author} {\bibfnamefont {J.}~\bibnamefont
  {Paier}}, \bibinfo {author} {\bibfnamefont {X.}~\bibnamefont {Ren}}, \bibinfo
  {author} {\bibfnamefont {P.}~\bibnamefont {Rinke}}, \bibinfo {author}
  {\bibfnamefont {G.~E.}\ \bibnamefont {Scuseria}}, \bibinfo {author}
  {\bibfnamefont {A.}~\bibnamefont {Gr{\"u}neis}}, \bibinfo {author}
  {\bibfnamefont {G.}~\bibnamefont {Kresse}},\ and\ \bibinfo {author}
  {\bibfnamefont {M.}~\bibnamefont {Scheffler}},\ }\href@noop {} {\bibfield
  {journal} {\bibinfo  {journal} {New J. Phys.}\ }\textbf {\bibinfo {volume}
  {14}},\ \bibinfo {pages} {043002} (\bibinfo {year} {2012})}\BibitemShut
  {NoStop}%
\bibitem [{\citenamefont {Ren}\ \emph {et~al.}(2013)\citenamefont {Ren},
  \citenamefont {Rinke}, \citenamefont {Scuseria},\ and\ \citenamefont
  {Scheffler}}]{Ren/etal:2013}%
  \BibitemOpen
  \bibfield  {author} {\bibinfo {author} {\bibfnamefont {X.}~\bibnamefont
  {Ren}}, \bibinfo {author} {\bibfnamefont {P.}~\bibnamefont {Rinke}}, \bibinfo
  {author} {\bibfnamefont {G.~E.}\ \bibnamefont {Scuseria}},\ and\ \bibinfo
  {author} {\bibfnamefont {M.}~\bibnamefont {Scheffler}},\ }\href@noop {}
  {\bibfield  {journal} {\bibinfo  {journal} {Phys. Rev. B}\ }\textbf {\bibinfo
  {volume} {88}},\ \bibinfo {pages} {035120} (\bibinfo {year}
  {2013})}\BibitemShut {NoStop}%
\bibitem [{\citenamefont {Ma}\ \emph {et~al.}(2019{\natexlab{a}})\citenamefont
  {Ma}, \citenamefont {Govoni}, \citenamefont {Gygi},\ and\ \citenamefont
  {Galli}}]{Ma/Galli:2019}%
  \BibitemOpen
  \bibfield  {author} {\bibinfo {author} {\bibfnamefont {H.}~\bibnamefont
  {Ma}}, \bibinfo {author} {\bibfnamefont {M.}~\bibnamefont {Govoni}}, \bibinfo
  {author} {\bibfnamefont {F.}~\bibnamefont {Gygi}},\ and\ \bibinfo {author}
  {\bibfnamefont {G.}~\bibnamefont {Galli}},\ }\href
  {https://doi.org/10.1021/acs.jctc.8b00864} {\bibfield  {journal} {\bibinfo
  {journal} {Journal of Chemical Theory and Computation}\ }\textbf {\bibinfo
  {volume} {15}},\ \bibinfo {pages} {154} (\bibinfo {year}
  {2019}{\natexlab{a}})},\ \Eprint
  {https://arxiv.org/abs/https://doi.org/10.1021/acs.jctc.8b00864}
  {https://doi.org/10.1021/acs.jctc.8b00864} \BibitemShut {NoStop}%
\bibitem [{\citenamefont {Uimonen}\ \emph {et~al.}(2015)\citenamefont
  {Uimonen}, \citenamefont {Stefanucci}, \citenamefont {Pavlyukh},\ and\
  \citenamefont {van Leeuwen}}]{Unimonen/vanLeeuwen:2014}%
  \BibitemOpen
  \bibfield  {author} {\bibinfo {author} {\bibfnamefont {A.-M.}\ \bibnamefont
  {Uimonen}}, \bibinfo {author} {\bibfnamefont {G.}~\bibnamefont {Stefanucci}},
  \bibinfo {author} {\bibfnamefont {Y.}~\bibnamefont {Pavlyukh}},\ and\
  \bibinfo {author} {\bibfnamefont {R.}~\bibnamefont {van Leeuwen}},\ }\href
  {https://doi.org/10.1103/PhysRevB.91.115104} {\bibfield  {journal} {\bibinfo
  {journal} {Phys. Rev. B}\ }\textbf {\bibinfo {volume} {91}},\ \bibinfo
  {pages} {115104} (\bibinfo {year} {2015})}\BibitemShut {NoStop}%
\bibitem [{\citenamefont {Stefanucci}\ \emph {et~al.}(2014)\citenamefont
  {Stefanucci}, \citenamefont {Pavlyukh}, \citenamefont {Uimonen},\ and\
  \citenamefont {van Leeuwen}}]{Stefanucci/vanLeeuwen:2015}%
  \BibitemOpen
  \bibfield  {author} {\bibinfo {author} {\bibfnamefont {G.}~\bibnamefont
  {Stefanucci}}, \bibinfo {author} {\bibfnamefont {Y.}~\bibnamefont
  {Pavlyukh}}, \bibinfo {author} {\bibfnamefont {A.-M.}\ \bibnamefont
  {Uimonen}},\ and\ \bibinfo {author} {\bibfnamefont {R.}~\bibnamefont {van
  Leeuwen}},\ }\href {https://doi.org/10.1103/PhysRevB.90.115134} {\bibfield
  {journal} {\bibinfo  {journal} {Phys. Rev. B}\ }\textbf {\bibinfo {volume}
  {90}},\ \bibinfo {pages} {115134} (\bibinfo {year} {2014})}\BibitemShut
  {NoStop}%
\bibitem [{\citenamefont {Pavlyukh}\ \emph
  {et~al.}(2020{\natexlab{b}})\citenamefont {Pavlyukh}, \citenamefont
  {Stefanucci},\ and\ \citenamefont {van Leeuwen}}]{Pavlyukh:vanLeeuwen:2020}%
  \BibitemOpen
  \bibfield  {author} {\bibinfo {author} {\bibfnamefont {Y.}~\bibnamefont
  {Pavlyukh}}, \bibinfo {author} {\bibfnamefont {G.}~\bibnamefont
  {Stefanucci}},\ and\ \bibinfo {author} {\bibfnamefont {R.}~\bibnamefont {van
  Leeuwen}},\ }\href {https://doi.org/10.1103/PhysRevB.102.045121} {\bibfield
  {journal} {\bibinfo  {journal} {Phys. Rev. B}\ }\textbf {\bibinfo {volume}
  {102}},\ \bibinfo {pages} {045121} (\bibinfo {year}
  {2020}{\natexlab{b}})}\BibitemShut {NoStop}%
\bibitem [{\citenamefont {Eshuis}\ \emph {et~al.}(2012)\citenamefont {Eshuis},
  \citenamefont {Bates},\ and\ \citenamefont
  {Furche}}]{Eshuis/Bates/Furche:2012}%
  \BibitemOpen
  \bibfield  {author} {\bibinfo {author} {\bibfnamefont {H.}~\bibnamefont
  {Eshuis}}, \bibinfo {author} {\bibfnamefont {J.~E.}\ \bibnamefont {Bates}},\
  and\ \bibinfo {author} {\bibfnamefont {F.}~\bibnamefont {Furche}},\
  }\href@noop {} {\bibfield  {journal} {\bibinfo  {journal} {Theor. Chem.
  Acc.}\ }\textbf {\bibinfo {volume} {131}},\ \bibinfo {pages} {1084} (\bibinfo
  {year} {2012})}\BibitemShut {NoStop}%
\bibitem [{\citenamefont {Blum}\ \emph {et~al.}(2009)\citenamefont {Blum},
  \citenamefont {Gehrke}, \citenamefont {Hanke}, \citenamefont {Havu},
  \citenamefont {Havu}, \citenamefont {Ren}, \citenamefont {Reuter},\ and\
  \citenamefont {Scheffler}}]{Blum2009}%
  \BibitemOpen
  \bibfield  {author} {\bibinfo {author} {\bibfnamefont {V.}~\bibnamefont
  {Blum}}, \bibinfo {author} {\bibfnamefont {R.}~\bibnamefont {Gehrke}},
  \bibinfo {author} {\bibfnamefont {F.}~\bibnamefont {Hanke}}, \bibinfo
  {author} {\bibfnamefont {P.}~\bibnamefont {Havu}}, \bibinfo {author}
  {\bibfnamefont {V.}~\bibnamefont {Havu}}, \bibinfo {author} {\bibfnamefont
  {X.}~\bibnamefont {Ren}}, \bibinfo {author} {\bibfnamefont {K.}~\bibnamefont
  {Reuter}},\ and\ \bibinfo {author} {\bibfnamefont {M.}~\bibnamefont
  {Scheffler}},\ }\href {https://doi.org/10.1016/j.cpc.2009.06.022} {\bibfield
  {journal} {\bibinfo  {journal} {Comput. Phys. Commun.}\ }\textbf {\bibinfo
  {volume} {180}},\ \bibinfo {pages} {2175} (\bibinfo {year}
  {2009})}\BibitemShut {NoStop}%
\bibitem [{\citenamefont {Havu}\ \emph {et~al.}(2009)\citenamefont {Havu},
  \citenamefont {Blum}, \citenamefont {Havu},\ and\ \citenamefont
  {Scheffler}}]{Havu/etal:2009}%
  \BibitemOpen
  \bibfield  {author} {\bibinfo {author} {\bibfnamefont {V.}~\bibnamefont
  {Havu}}, \bibinfo {author} {\bibfnamefont {V.}~\bibnamefont {Blum}}, \bibinfo
  {author} {\bibfnamefont {P.}~\bibnamefont {Havu}},\ and\ \bibinfo {author}
  {\bibfnamefont {M.}~\bibnamefont {Scheffler}},\ }\href@noop {} {\bibfield
  {journal} {\bibinfo  {journal} {J. Comp. Phys.}\ }\textbf {\bibinfo {volume}
  {228}},\ \bibinfo {pages} {8367} (\bibinfo {year} {2009})}\BibitemShut
  {NoStop}%
\bibitem [{\citenamefont {Dunlap}\ \emph {et~al.}(1979)\citenamefont {Dunlap},
  \citenamefont {Connolly},\ and\ \citenamefont
  {Sabin}}]{Dunlap/Connolly/Sabin:1979}%
  \BibitemOpen
  \bibfield  {author} {\bibinfo {author} {\bibfnamefont {B.~I.}\ \bibnamefont
  {Dunlap}}, \bibinfo {author} {\bibfnamefont {J.~W.~D.}\ \bibnamefont
  {Connolly}},\ and\ \bibinfo {author} {\bibfnamefont {J.~R.}\ \bibnamefont
  {Sabin}},\ }\href@noop {} {\bibfield  {journal} {\bibinfo  {journal} {J.
  Chem. Phys}\ }\textbf {\bibinfo {volume} {71}},\ \bibinfo {pages} {3396}
  (\bibinfo {year} {1979})}\BibitemShut {NoStop}%
\bibitem [{\citenamefont {Whitten}(1973)}]{Whitten:1973}%
  \BibitemOpen
  \bibfield  {author} {\bibinfo {author} {\bibfnamefont {J.~L.}\ \bibnamefont
  {Whitten}},\ }\href@noop {} {\bibfield  {journal} {\bibinfo  {journal} {J.
  Chem. Phys.}\ }\textbf {\bibinfo {volume} {58}},\ \bibinfo {pages} {4496}
  (\bibinfo {year} {1973})}\BibitemShut {NoStop}%
\bibitem [{\citenamefont {Vahtras}\ \emph {et~al.}(1993)\citenamefont
  {Vahtras}, \citenamefont {Alml{\"o}f},\ and\ \citenamefont
  {Feyereisen}}]{Vahtras/Almlof/Feyereisen:1993}%
  \BibitemOpen
  \bibfield  {author} {\bibinfo {author} {\bibfnamefont {O.}~\bibnamefont
  {Vahtras}}, \bibinfo {author} {\bibfnamefont {J.}~\bibnamefont
  {Alml{\"o}f}},\ and\ \bibinfo {author} {\bibfnamefont {M.~W.}\ \bibnamefont
  {Feyereisen}},\ }\href@noop {} {\bibfield  {journal} {\bibinfo  {journal}
  {Chem. Phys. Lett.}\ }\textbf {\bibinfo {volume} {213}},\ \bibinfo {pages}
  {514} (\bibinfo {year} {1993})}\BibitemShut {NoStop}%
\bibitem [{\citenamefont {Baker}(1975)}]{Baker1975}%
  \BibitemOpen
  \bibfield  {author} {\bibinfo {author} {\bibfnamefont {G.~A.}\ \bibnamefont
  {Baker}},\ }in\ \href@noop {} {\emph {\bibinfo {booktitle} {Essentials of
  Pad$\acute{\text{e}}$ Approximants}}}\ (\bibinfo  {publisher} {Academic
  Press},\ \bibinfo {address} {New York},\ \bibinfo {year} {1975})\
  Chap.~\bibinfo {chapter} {8}, p.\ \bibinfo {pages} {105}\BibitemShut
  {NoStop}%
\bibitem [{\citenamefont {T.~H.~Dunning}(1989)}]{Dunning:1989}%
  \BibitemOpen
  \bibfield  {author} {\bibinfo {author} {\bibfnamefont {J.}~\bibnamefont
  {T.~H.~Dunning}},\ }\href@noop {} {\bibfield  {journal} {\bibinfo  {journal}
  {J. Chem. Phys.}\ }\textbf {\bibinfo {volume} {90}},\ \bibinfo {pages} {1007}
  (\bibinfo {year} {1989})}\BibitemShut {NoStop}%
\bibitem [{\citenamefont {Richard}\ \emph {et~al.}(2016)\citenamefont
  {Richard}, \citenamefont {Marshall}, \citenamefont {Dolgounitcheva},
  \citenamefont {Ortiz}, \citenamefont {Brédas}, \citenamefont {Marom},\ and\
  \citenamefont {Sherrill}}]{Richard2016/10.1021/acs.jctc.5b00875}%
  \BibitemOpen
  \bibfield  {author} {\bibinfo {author} {\bibfnamefont {R.~M.}\ \bibnamefont
  {Richard}}, \bibinfo {author} {\bibfnamefont {M.~S.}\ \bibnamefont
  {Marshall}}, \bibinfo {author} {\bibfnamefont {O.}~\bibnamefont
  {Dolgounitcheva}}, \bibinfo {author} {\bibfnamefont {J.~V.}\ \bibnamefont
  {Ortiz}}, \bibinfo {author} {\bibfnamefont {J.-L.}\ \bibnamefont {Brédas}},
  \bibinfo {author} {\bibfnamefont {N.}~\bibnamefont {Marom}},\ and\ \bibinfo
  {author} {\bibfnamefont {C.~D.}\ \bibnamefont {Sherrill}},\ }\href
  {https://doi.org/10.1021/acs.jctc.5b00875} {\bibfield  {journal} {\bibinfo
  {journal} {J. Chem. Theory Comput.}\ }\textbf {\bibinfo {volume} {12}},\
  \bibinfo {pages} {595} (\bibinfo {year} {2016})}\BibitemShut {NoStop}%
\bibitem [{\citenamefont {Katharina~Krause}\ and\ \citenamefont
  {Klopper}(2015)}]{Katharina/Klopper:2015}%
  \BibitemOpen
  \bibfield  {author} {\bibinfo {author} {\bibfnamefont {M.~E.~H.}\
  \bibnamefont {Katharina~Krause}}\ and\ \bibinfo {author} {\bibfnamefont
  {W.}~\bibnamefont {Klopper}},\ }\href
  {https://doi.org/10.1080/00268976.2015.1025113} {\bibfield  {journal}
  {\bibinfo  {journal} {Mol. Phys.}\ }\textbf {\bibinfo {volume} {113}},\
  \bibinfo {pages} {1952} (\bibinfo {year} {2015})}\BibitemShut {NoStop}%
\bibitem [{\citenamefont {Perdew}\ \emph
  {et~al.}(1996{\natexlab{a}})\citenamefont {Perdew}, \citenamefont {Burke},\
  and\ \citenamefont {Ernzerhof}}]{Perdew/Burke/Ernzerhof:1996}%
  \BibitemOpen
  \bibfield  {author} {\bibinfo {author} {\bibfnamefont {J.~P.}\ \bibnamefont
  {Perdew}}, \bibinfo {author} {\bibfnamefont {K.}~\bibnamefont {Burke}},\ and\
  \bibinfo {author} {\bibfnamefont {M.}~\bibnamefont {Ernzerhof}},\ }\href
  {https://doi.org/10.1103/PhysRevLett.77.3865} {\bibfield  {journal} {\bibinfo
   {journal} {Phys. Rev. Lett.}\ }\textbf {\bibinfo {volume} {77}},\ \bibinfo
  {pages} {3865} (\bibinfo {year} {1996}{\natexlab{a}})}\BibitemShut {NoStop}%
\bibitem [{\citenamefont {Perdew}\ \emph
  {et~al.}(1996{\natexlab{b}})\citenamefont {Perdew}, \citenamefont
  {Ernzerhof},\ and\ \citenamefont {Burke}}]{Perdew/Ernzerhof/Burke:1996}%
  \BibitemOpen
  \bibfield  {author} {\bibinfo {author} {\bibfnamefont {J.~P.}\ \bibnamefont
  {Perdew}}, \bibinfo {author} {\bibfnamefont {M.}~\bibnamefont {Ernzerhof}},\
  and\ \bibinfo {author} {\bibfnamefont {K.}~\bibnamefont {Burke}},\
  }\href@noop {} {\bibfield  {journal} {\bibinfo  {journal} {J. Chem. Phys.}\
  }\textbf {\bibinfo {volume} {105}},\ \bibinfo {pages} {9982} (\bibinfo {year}
  {1996}{\natexlab{b}})}\BibitemShut {NoStop}%
\bibitem [{\citenamefont {Adamo}\ and\ \citenamefont
  {Barone}(1999)}]{Carlo1999}%
  \BibitemOpen
  \bibfield  {author} {\bibinfo {author} {\bibfnamefont {C.}~\bibnamefont
  {Adamo}}\ and\ \bibinfo {author} {\bibfnamefont {V.}~\bibnamefont {Barone}},\
  }\href {https://doi.org/10.1063/1.478522} {\bibfield  {journal} {\bibinfo
  {journal} {J. Chem. Phys.}\ }\textbf {\bibinfo {volume} {110}},\ \bibinfo
  {pages} {6158} (\bibinfo {year} {1999})}\BibitemShut {NoStop}%
\bibitem [{\citenamefont {Weigenda}\ and\ \citenamefont
  {Ahlrichsb}(2005)}]{Florian2005}%
  \BibitemOpen
  \bibfield  {author} {\bibinfo {author} {\bibfnamefont {F.}~\bibnamefont
  {Weigenda}}\ and\ \bibinfo {author} {\bibfnamefont {R.}~\bibnamefont
  {Ahlrichsb}},\ }\href {https://doi.org/10.1039/B508541A} {\bibfield
  {journal} {\bibinfo  {journal} {Phys. Chem. Chem. Phys.}\ }\textbf {\bibinfo
  {volume} {7}},\ \bibinfo {pages} {3297} (\bibinfo {year} {2005})}\BibitemShut
  {NoStop}%
\bibitem [{\citenamefont {Rangel}\ \emph {et~al.}(2016)\citenamefont {Rangel},
  \citenamefont {Hamed}, \citenamefont {Bruneval},\ and\ \citenamefont
  {Neaton}}]{Rangel2016}%
  \BibitemOpen
  \bibfield  {author} {\bibinfo {author} {\bibfnamefont {T.}~\bibnamefont
  {Rangel}}, \bibinfo {author} {\bibfnamefont {S.~M.}\ \bibnamefont {Hamed}},
  \bibinfo {author} {\bibfnamefont {F.}~\bibnamefont {Bruneval}},\ and\
  \bibinfo {author} {\bibfnamefont {J.~B.}\ \bibnamefont {Neaton}},\ }\href
  {https://doi.org/10.1021/acs.jctc.6b00163} {\bibfield  {journal} {\bibinfo
  {journal} {J. Chem. Theory Comput.}\ }\textbf {\bibinfo {volume} {12}},\
  \bibinfo {pages} {2834} (\bibinfo {year} {2016})}\BibitemShut {NoStop}%
\bibitem [{\citenamefont {Gunnarsson}\ \emph {et~al.}(2010)\citenamefont
  {Gunnarsson}, \citenamefont {Haverkort},\ and\ \citenamefont
  {Sangiovanni}}]{Gunnarsson2010}%
  \BibitemOpen
  \bibfield  {author} {\bibinfo {author} {\bibfnamefont {O.}~\bibnamefont
  {Gunnarsson}}, \bibinfo {author} {\bibfnamefont {M.~W.}\ \bibnamefont
  {Haverkort}},\ and\ \bibinfo {author} {\bibfnamefont {G.}~\bibnamefont
  {Sangiovanni}},\ }\href {https://doi.org/10.1103/PhysRevB.82.165125}
  {\bibfield  {journal} {\bibinfo  {journal} {Phys. Rev. B}\ }\textbf {\bibinfo
  {volume} {82}},\ \bibinfo {pages} {165125} (\bibinfo {year}
  {2010})}\BibitemShut {NoStop}%
\bibitem [{\citenamefont {Duchemin}\ and\ \citenamefont
  {Blase}(2020)}]{Blase2020}%
  \BibitemOpen
  \bibfield  {author} {\bibinfo {author} {\bibfnamefont {I.}~\bibnamefont
  {Duchemin}}\ and\ \bibinfo {author} {\bibfnamefont {X.}~\bibnamefont
  {Blase}},\ }\href {https://doi.org/10.1021/acs.jctc.9b01235} {\bibfield
  {journal} {\bibinfo  {journal} {J. Chem. Theory Comput.}\ }\textbf {\bibinfo
  {volume} {16}},\ \bibinfo {pages} {1742} (\bibinfo {year}
  {2020})}\BibitemShut {NoStop}%
\bibitem [{\citenamefont {Deslippe}\ \emph {et~al.}(2012)\citenamefont
  {Deslippe}, \citenamefont {Samsonidze}, \citenamefont {Strubbe},
  \citenamefont {Jain}, \citenamefont {L.Cohen},\ and\ \citenamefont
  {G.Louie}}]{Deslippe/etal:2012}%
  \BibitemOpen
  \bibfield  {author} {\bibinfo {author} {\bibfnamefont {J.}~\bibnamefont
  {Deslippe}}, \bibinfo {author} {\bibfnamefont {G.}~\bibnamefont
  {Samsonidze}}, \bibinfo {author} {\bibfnamefont {D.~A.}\ \bibnamefont
  {Strubbe}}, \bibinfo {author} {\bibfnamefont {M.}~\bibnamefont {Jain}},
  \bibinfo {author} {\bibfnamefont {M.}~\bibnamefont {L.Cohen}},\ and\ \bibinfo
  {author} {\bibfnamefont {S.}~\bibnamefont {G.Louie}},\ }\href@noop {}
  {\bibfield  {journal} {\bibinfo  {journal} {Comput. Phys. Commun.}\ }\textbf
  {\bibinfo {volume} {183}},\ \bibinfo {pages} {1269} (\bibinfo {year}
  {2012})}\BibitemShut {NoStop}%
\bibitem [{\citenamefont {Ma}\ \emph {et~al.}(2019{\natexlab{b}})\citenamefont
  {Ma}, \citenamefont {Govoni}, \citenamefont {Gygi},\ and\ \citenamefont
  {Galli}}]{Ma/Galli/etal:2019}%
  \BibitemOpen
  \bibfield  {author} {\bibinfo {author} {\bibfnamefont {H.}~\bibnamefont
  {Ma}}, \bibinfo {author} {\bibfnamefont {M.}~\bibnamefont {Govoni}}, \bibinfo
  {author} {\bibfnamefont {F.}~\bibnamefont {Gygi}},\ and\ \bibinfo {author}
  {\bibfnamefont {G.}~\bibnamefont {Galli}},\ }\href
  {https://doi.org/10.1021/acs.jctc.8b00864} {\bibfield  {journal} {\bibinfo
  {journal} {J. Chem. Theory Comput.}\ }\textbf {\bibinfo {volume} {15}},\
  \bibinfo {pages} {154} (\bibinfo {year} {2019}{\natexlab{b}})}\BibitemShut
  {NoStop}%
\bibitem [{GW1()}]{GW100press}%
  \BibitemOpen
  \href@noop {} {}\bibinfo {note}
  {\url{http://gw100.worldpress.com/}}\BibitemShut {NoStop}%
\bibitem [{\citenamefont {Marom}(2017)}]{Marom:2017}%
  \BibitemOpen
  \bibfield  {author} {\bibinfo {author} {\bibfnamefont {N.}~\bibnamefont
  {Marom}},\ }\href {https://doi.org/10.1088/1361-648x/29/10/103003} {\bibfield
   {journal} {\bibinfo  {journal} {J. Condens. Matter Phys.}\ }\textbf
  {\bibinfo {volume} {29}},\ \bibinfo {pages} {103003} (\bibinfo {year}
  {2017})}\BibitemShut {NoStop}%
\bibitem [{\citenamefont {Gallandi}\ \emph {et~al.}(2016)\citenamefont
  {Gallandi}, \citenamefont {Marom}, \citenamefont {Rinke},\ and\ \citenamefont
  {Körzdörfer}}]{Gallandi/etal:2016}%
  \BibitemOpen
  \bibfield  {author} {\bibinfo {author} {\bibfnamefont {L.}~\bibnamefont
  {Gallandi}}, \bibinfo {author} {\bibfnamefont {N.}~\bibnamefont {Marom}},
  \bibinfo {author} {\bibfnamefont {P.}~\bibnamefont {Rinke}},\ and\ \bibinfo
  {author} {\bibfnamefont {T.}~\bibnamefont {Körzdörfer}},\ }\href
  {https://doi.org/10.1021/acs.jctc.5b00873} {\bibfield  {journal} {\bibinfo
  {journal} {Journal of Chemical Theory and Computation}\ }\textbf {\bibinfo
  {volume} {12}},\ \bibinfo {pages} {605} (\bibinfo {year} {2016})},\ \bibinfo
  {note} {pMID: 26731340}\BibitemShut {NoStop}%
\bibitem [{\citenamefont {Dolgounitcheva}\ \emph {et~al.}(2016)\citenamefont
  {Dolgounitcheva}, \citenamefont {Díaz-Tinoco}, \citenamefont {Zakrzewski},
  \citenamefont {Richard}, \citenamefont {Marom}, \citenamefont {Sherrill},\
  and\ \citenamefont {Ortiz}}]{Dolgounitcheva/etal:2016}%
  \BibitemOpen
  \bibfield  {author} {\bibinfo {author} {\bibfnamefont {O.}~\bibnamefont
  {Dolgounitcheva}}, \bibinfo {author} {\bibfnamefont {M.}~\bibnamefont
  {Díaz-Tinoco}}, \bibinfo {author} {\bibfnamefont {V.~G.}\ \bibnamefont
  {Zakrzewski}}, \bibinfo {author} {\bibfnamefont {R.~M.}\ \bibnamefont
  {Richard}}, \bibinfo {author} {\bibfnamefont {N.}~\bibnamefont {Marom}},
  \bibinfo {author} {\bibfnamefont {C.~D.}\ \bibnamefont {Sherrill}},\ and\
  \bibinfo {author} {\bibfnamefont {J.~V.}\ \bibnamefont {Ortiz}},\ }\href
  {https://doi.org/10.1021/acs.jctc.5b00872} {\bibfield  {journal} {\bibinfo
  {journal} {Journal of Chemical Theory and Computation}\ }\textbf {\bibinfo
  {volume} {12}},\ \bibinfo {pages} {627} (\bibinfo {year} {2016})},\ \bibinfo
  {note} {pMID: 26730459}\BibitemShut {NoStop}%
\bibitem [{\citenamefont {Liu}\ \emph {et~al.}(2011)\citenamefont {Liu},
  \citenamefont {Alnama}, \citenamefont {Matsumoto}, \citenamefont {Nishizawa},
  \citenamefont {Kohguchi}, \citenamefont {Lee}, ,\ and\ \citenamefont
  {Suzuki}}]{Liu/etal:2011}%
  \BibitemOpen
  \bibfield  {author} {\bibinfo {author} {\bibfnamefont {S.-Y.}\ \bibnamefont
  {Liu}}, \bibinfo {author} {\bibfnamefont {K.}~\bibnamefont {Alnama}},
  \bibinfo {author} {\bibfnamefont {J.}~\bibnamefont {Matsumoto}}, \bibinfo
  {author} {\bibfnamefont {K.}~\bibnamefont {Nishizawa}}, \bibinfo {author}
  {\bibfnamefont {H.}~\bibnamefont {Kohguchi}}, \bibinfo {author}
  {\bibfnamefont {Y.-P.}\ \bibnamefont {Lee}}, ,\ and\ \bibinfo {author}
  {\bibfnamefont {T.}~\bibnamefont {Suzuki}},\ }\href@noop {} {\bibfield
  {journal} {\bibinfo  {journal} {J. Phys. Chem. A}\ }\textbf {\bibinfo
  {volume} {115}},\ \bibinfo {pages} {2953} (\bibinfo {year}
  {2011})}\BibitemShut {NoStop}%
\bibitem [{\citenamefont {Jin}\ \emph {et~al.}(2019)\citenamefont {Jin},
  \citenamefont {Su},\ and\ \citenamefont {Yang}}]{Jin/Su/Yang:2019}%
  \BibitemOpen
  \bibfield  {author} {\bibinfo {author} {\bibfnamefont {Y.}~\bibnamefont
  {Jin}}, \bibinfo {author} {\bibfnamefont {N.~Q.}\ \bibnamefont {Su}},\ and\
  \bibinfo {author} {\bibfnamefont {W.}~\bibnamefont {Yang}},\ }\href
  {https://doi.org/10.1021/acs.jpclett.8b03337} {\bibfield  {journal} {\bibinfo
   {journal} {J. Phys. Chem. Lett.}\ }\textbf {\bibinfo {volume} {10}},\
  \bibinfo {pages} {447} (\bibinfo {year} {2019})}\BibitemShut {NoStop}%
\bibitem [{\citenamefont {Ranasinghe}\ \emph {et~al.}(2019)\citenamefont
  {Ranasinghe}, \citenamefont {Margraf}, \citenamefont {Perera},\ and\
  \citenamefont {Bartlett}}]{Ranasinghe/etal:2019}%
  \BibitemOpen
  \bibfield  {author} {\bibinfo {author} {\bibfnamefont {D.~S.}\ \bibnamefont
  {Ranasinghe}}, \bibinfo {author} {\bibfnamefont {J.~T.}\ \bibnamefont
  {Margraf}}, \bibinfo {author} {\bibfnamefont {A.}~\bibnamefont {Perera}},\
  and\ \bibinfo {author} {\bibfnamefont {R.~J.}\ \bibnamefont {Bartlett}},\
  }\href {https://doi.org/10.1063/1.5084728} {\bibfield  {journal} {\bibinfo
  {journal} {The Journal of Chemical Physics}\ }\textbf {\bibinfo {volume}
  {150}},\ \bibinfo {pages} {074108} (\bibinfo {year} {2019})}\BibitemShut
  {NoStop}%
\bibitem [{\citenamefont {Baym}\ and\ \citenamefont
  {Kadanoff}(1961)}]{Baym/Kadanoff:1961}%
  \BibitemOpen
  \bibfield  {author} {\bibinfo {author} {\bibfnamefont {G.}~\bibnamefont
  {Baym}}\ and\ \bibinfo {author} {\bibfnamefont {L.~P.}\ \bibnamefont
  {Kadanoff}},\ }\href@noop {} {\bibfield  {journal} {\bibinfo  {journal}
  {Phys. Rev.}\ }\textbf {\bibinfo {volume} {124}},\ \bibinfo {pages} {287}
  (\bibinfo {year} {1961})}\BibitemShut {NoStop}%
\bibitem [{\citenamefont {Almbladh}\ \emph {et~al.}(1999)\citenamefont
  {Almbladh}, \citenamefont {{von Barth}},\ and\ \citenamefont {{van
  Leeuwen}}}]{Almbladh/etal:1999}%
  \BibitemOpen
  \bibfield  {author} {\bibinfo {author} {\bibfnamefont {C.-O.}\ \bibnamefont
  {Almbladh}}, \bibinfo {author} {\bibfnamefont {U.}~\bibnamefont {{von
  Barth}}},\ and\ \bibinfo {author} {\bibfnamefont {R.}~\bibnamefont {{van
  Leeuwen}}},\ }\href@noop {} {\bibfield  {journal} {\bibinfo  {journal} {Int.
  J. Mod. Phys. B}\ }\textbf {\bibinfo {volume} {13}},\ \bibinfo {pages} {535}
  (\bibinfo {year} {1999})}\BibitemShut {NoStop}%
\bibitem [{\citenamefont {Dahlen}\ \emph {et~al.}(2006)\citenamefont {Dahlen},
  \citenamefont {{van Leeuwen}},\ and\ \citenamefont {{von
  Barth}}}]{Dahlen/Leeuwen/Barth:2006}%
  \BibitemOpen
  \bibfield  {author} {\bibinfo {author} {\bibfnamefont {N.~E.}\ \bibnamefont
  {Dahlen}}, \bibinfo {author} {\bibfnamefont {R.}~\bibnamefont {{van
  Leeuwen}}},\ and\ \bibinfo {author} {\bibfnamefont {U.}~\bibnamefont {{von
  Barth}}},\ }\href@noop {} {\bibfield  {journal} {\bibinfo  {journal} {Phys.
  Rev. A}\ }\textbf {\bibinfo {volume} {73}},\ \bibinfo {pages} {012511}
  (\bibinfo {year} {2006})}\BibitemShut {NoStop}%
\end{thebibliography}%
\end{document}